\documentclass[11pt]{article} % Specifies font size
\usepackage[utf8]{inputenc}
\usepackage{libertine}
\usepackage[left=2cm,right=2cm,top=2cm,bottom=2cm]{geometry}
\usepackage[pdftex]{graphicx} % Allows inclusion of image files
\usepackage{amssymb} % Access to extra math symbols
\usepackage{amsmath} % Access to extra math symbols
\usepackage{amsthm}
\usepackage{wrapfig} % Allows wrapping of text around figures
\usepackage{calc} % Gives access to a basic calculator 
\usepackage{footmisc} % footnotes better
\usepackage{empheq}
\usepackage{amsmath}
\usepackage[table, svgnames, dvipsnames]{xcolor}
\usepackage{makecell, cellspace, caption}
\usepackage{subcaption}
\usepackage{lineno}
\usepackage{mathtools}
\usepackage{url}
\usepackage{hyperref}
\usepackage{appendix}
\hypersetup{colorlinks,linkcolor={},citecolor={blue},urlcolor={red}}

\usepackage[version=4]{mhchem}

\usepackage{booktabs}
\usepackage[square,sort,comma,numbers]{natbib}
\usepackage{epigraph}
\usepackage{etoolbox}
\usepackage{authblk}
\makeatletter

\patchcmd{\maketitle}{\@fnsymbol}{\@alph}{}{}
\makeatother

\title{Incremental formation of scale-free fitness networks}

\author[a,b]{Fabio Vanni\footnote{Correspondent author: fabio.vanni@sciencespo.fr }}

\affil[a]{Sciences Po, OFCE , France}
\affil[b]{Center for Nonlinear Sciences, University of North Texas, USA}

\date{}

\begin{document}

\maketitle

% Abstract; JEL codes, and keywords
\begin{abstract}
In the  paper, we  present an incremental approach in the construction of scale free networks with hidden variables. The  work arises from the necessity to  generate that type of networks with a given number of links instead of obtaining a random configurations for a given set of parameters as in the usual literature. 
I propose an analytical approach of  network  evolution  models  gathering  information  along  time based   on  the  construction  of  a  stochastic  process  on  the  space  of  possible  networks. The analytical solution is eact in a case of multigraph network, meanwhile in simple graph we deal with important finite size effects. 
We show the statistical properties of this network such as number of isolated nodes, degree correlations and multilinks, also discussing  the limitations of such predictions in real networks.  Numerical simulations are used tu support the analytical calculations.
On the computational side, such evolutive network construction allows to regulate the connectivity of the network obtaining  desired values of  connectivity density used as control parameter of the system.  As a consequence, an incremental generetive approach is more suitable in those situation we want to control the number of links in the system while the number of nodes are kept fixed.
\end{abstract}

% % % % % % % % % % % %        SECTION 1          % % % % % % % % % % % % % % % % % % % % % % % % % % % % % % % % % % % % % %
\section{Introduction}\label{sec:intro}
Network theory provides a natural way to model interconnections and its application to banking and finance has brought to a growing body of academic research on financial networks, bloomed over the past few years. The network approach to risk and stability has challenged the standard wisdom that a higher degree of interconnectedness is always desirable to achieve financial stability.  A large part of network theory to financial markets has been focused on the analysis on how unexpected exogenous shocks propagate through a complex financial network where shocks can force a financial units to default.
Diverse authors have developed a contagion  and systemic risk models with both random ans scale-free structure networks as by \citet{gai2010contagion} and \cite{montagna2013hubs}.  
%In this paper we provide a general framework to assess the risk of a system wide crisis and the resilience of financial networks. We present a model with two types of agents, banks and firms, linked one another in a two layer structure by their reciprocal claims. The model proposed is able to assess the frequency and the extent of a systemic crisis and to track its dynamical evolution. %, as well as to gauge the resilience of the system to shocks of different intensity. 
%We simulate the model numerically and compare the result for different levels of interconnection of the network, degrees of heterogeneity and types of shock.
%
% Accordingly, the numerical results in terms of frequency and extent of crisis do not claim to be neither accurate nor rough estimates of real world dynamics or to constitute a basis for any forecast. Despite this cautions, the present work provides an interesting insight for the emergence of a systemic crisis and represents a first attempt to link together the financial and the real side of the economy, two aspects too often treated separately, constituting a useful starting point for future research.
In the following paper we  describe a class of network models with attachment rules governed by intrinsic node fitness. This fitness is a measure of attractiveness and also a measure of a node's robustness against failure.  In the network representation agents are the nodes of the graph while the edges encode interactions. In the field of networks formation literature \citet{doronetwork}, there are random generated models where the set of nodes are fixed and dynamic networks growth models where or number of nodes and/or edges changes over time. In the present work we deal with a random generated model for scale-free networks where, contrary to the standard approach \citet{caldarelli2016data}, it is possible to control the total number of edges (equivalently the average degree) at the end of procedure.  To some extent, our prescription is the equivalent of Erdos-Renyi random graph generation as opposite to the Gilbert prescription. Despite the fact that the two prescriptions bring to the same degree distribution in  limit of infinite number of nodes. 
Hidden variable models are models that avoid the need of assumptions of perfect knowledge about network topology (as preferential attachment).  The linking rules for edges in such models are governed by some measure of the intrinsic quality of the node.  In fact, hidden variable models are purely driven by static node intrinsic fitness and the topology of the network is determined by an attachment kernel function that describes the probability that a node with a certain fitness originates a link towards a node with another fitness, and the probability density that describes the distribution of fitness in the system. On the other side, degree dependent growth models new nodes have information about the connectivity of the entire network (Barabasi's preferential attachment).  For fitness driven networks it is sufficient to assume that nodes have information about the ranking of some derived topological independent quantity. For example, the balance sheet structure of a firm or a bank is more likely to be accessible to an investor than the absolute number of other investors invested in a particular company.  
The association of attractiveness and robustness to a failure is of great interest to organizations that can influence the underlying free parameters, e.g. distribution of fitness, as well as the average node-degree in the system as in in the interbank lending market, where the linking function is given fixed and cannot be influenced, but policy maker or regulator can reshape the fitness distribution by the introduction of a tax. 

A  graph  process, denoted  by $G$  is  a  method  of  constructing  a network by adding a link to a graph $G_0$ by choosing one with equal likelihood out of all possible links. $G_t$ is the network constructed after $t$ of these steps. 
The characterization  of  complex  networks  can be  addressed  by  considering  them  as  either  static  or  dynamic  entities.  The  most  frequent  static  perspective  is  grounded  on  constructing  a  probability  space  (i.e.,  a  probability  measure  on  the  space  of  possible  networks);  this  probability  model  allows  for  a  compact  network  characterization. Network  evolution  models  gather  information  along  time  and  they  are   grounded  on  the  construction  of  a  stochastic  process  on  the  space  of  possible  networks. 
In particular, the fitness network model can be achieved following two approaches:
 \begin{enumerate}
 \item \textit{Random Generation $G(N,p)$}: is a probability space on the setof graphs with $N$ nodes where each link in the graph is added independently with probability $p$. The number of links $L$ is a random variable. We calculate once the matrix of making a link per each node and then assign randomly $1$'s and $0$'s to the adjacency matrix of the network. 
 \item \textit{Incremental Generation $G(N,L)$}: is a probability space of all unlabeled graphs with $N$ nodes and $L$ links. We start with $N$ isolated nodes and connect two nodes per time  step. The number of links $L$ is chosen to be fixed and deterministic. In this case we can control the average degree $\langle k \rangle$ adding $L$ links to the network one by one.
 \end{enumerate}
 In both versions,  fitness models are characterized  by an attachment kernel  and a fitness distribution $\rho(x)$. The attachment kernel $f(x, y)$  describes the probability that a node with fitness $x$ originates a new edge  toward a node with fitness $y$. The fitness of each node is
 static over the lifetime of the network and is drawn from
 the probability density $\rho(x)$. 
The classical random approach to the study of hidden variable graph rests on the work of \citet{caldarelli2002scale,servedio2004vertex} who find directly the degree distribution and some statistical properties. A generalization of such models has been deeply addressed by \citet{boguna2003class,serrano2007large} which makes uses of a microscopic description of the system. Finally a more recent approach instead, look at such network from a evolutionary perspective, where links are sequentially deployed at each time step \citet{bedogne2006complex,hoppe2015microscopic}.   
 The paper contributes to give an analytical description of a incremental formation of hidden variable networks with factorizable  linking probability. 
 We provide a master equation for the evolution of  the degree probability distribution, and the evolution of expected number of isolated nodes.
Further we will discuss some properties of the networks such as its degree distribution shape over time and its finite size effects as well as correlations among nodes in the case both of simple graph and multi-link version of the model.
At the end, in the paper,  we  compare the previous analytical estimations to the Montecarlo simulations of the model to test the reliability of the theoretical description  so giving the possibility to discuss some important implications on the use of a incremental formulation  of hidden variable networks. 
 
 \section{Incremental Model}
   Among all the possible real applications, in financial situations it is  reasonable to think that two nodes will be connected depending on some of their intrinsic properties such as social status or  information content. In this way, two nodes become connected when the link creates a mutual benefit \citet{caldarelli2007scale}.
  Therefore we build a financial network model in which each node $i$ $(i = 1, . . . , N )$ is  assigned a " fitness " described by a random real variable $x_i$ distributed according to  a certain probability distribution $\rho(x)$.  Each pair of nodes $(i, j)$ is then connected with a probability depending on the fitnesses or types of the two nodes, $p_{i,j} = f (x_i , x_j )$, where $f$ is a given function\footnote{In the case where $f = const.$, the present model is equivalent to an Erdos-Renyi  graph (random graph).}. 
  In this model the probability to have a link between two nodes depend on the fitness of the two nodes via a bi-dimensional  linking function. 
  There are different particular choices of such function, as expressed in \citet{caldarelli2007scale} and \citet{montagna2013hubs}. The one we use in the present paper is a factorizable form of the type:
  \begin{equation}
  f(x,y)=g(x)h(y),
  \end{equation}
  where in principle $x$ and $y$ can be two different peculiarities of the node, for example, in stock network,  one peculiarity of the node can be the portfolio volume and the other the asset information. In the  case of bank financial market  such function provides the probability that a bank $i$ lends money to bank $j$, and such banks (nodes) have only one  peculiarity that is their financial size (assets for example) $x_i$ and $x_j$\footnote{Note that using the terminology of DEBIT/CREDIT the linking probability are reversed in terms of out-degree and in-degree, because the links are reversed in directions: who lands is creditor, who borrows is debtor. Now  the links goes from debtors to creditors because debtors should give back the money they borrowed. so $f(x_i,x_j)=\tilde{f}(x_j,x_i)$, where $\tilde{f}$ is the linking function for debit/credit connection so in the following formulas we nedd to exchange the term $in$ to $out$ if we speaks in terms of credits and debits for all network properties. In other terms the Adjacency matrix in trasposed (or just exchange the exponents $\alpha$ and $\beta$)}.  Such peculiarites can be consider as random variable drawn from a density function. Combination of different fitness distributions and linking functions can bring to different structures and behaviors of networks.  
  The empirical observations int he field of interbank markets suggest an appropriate linking function that reproduce the fact that small banks usually lend money to the biggest banks of the system, which in turn redistribute liquidity to the external financial markets or within the interbank market itself.
As regard with the fitness $\rho (x)$, one can choose any kind of distribution, however we select a Pareto distribution, particularly in its truncated version as widely discussed in the next sections.   
At this point we introduce the temporal prescription to generate a network that describes a generalization of the hidden variable models for a incremental point of view. Anyway we will still recover the particular situation in asymptotic regime.
This kind of procedure has also another technical application (as in \citet{BODEVA2016}), it allows to generate a given type of scale-free fitness network with a given connectivity (or average degree for finite networks). 
 The network building algorithm is derived with the following steps:
 \begin{itemize}
 \item Start with $N$ nodes. Links are  deployed one by one, not all at the same time.
 
 \item A link is added to a pair of nodes $(i , j)$ with probability,
 $f (x_i , x_j )/ \sum_{k,l} f (x_k , x_l )$.  from $j$ to $i$ in a directed outgoing direction.
 
 \item The processes of adding nodes continues until a given number of links $L$ have been added.
 
 \end{itemize}
 
 In the present model the number of nodes remains constant during the network's evolution, while the number of links increases linearly with time. As a result  the average degree  also increases  linearly with time:
 \begin{equation}
 \langle {k}(t)\rangle \sim \frac{t}{N} =  \frac{L}{N} 
 \end{equation}
 In the simple graph formation, at each time step, each new link connects previously unconnected nodes, so that the number of links coincides with time passed from th origin.. After a sufficient number of links have been created the graph go towards the all to all condition for $L \to N(N-1)/2$ so that the network becomes a complete   graph in which all nodes have degree $k_{max}=N-1$ and even if the time passes no new link can be created, so the correspondence between number of links and time breaks.
Although the network is static i.e. it contains a fixed number of nodes, the process in which edges are added can be understood as a dynamic procedure. Links are deployed one by one, not all at the same time snapshot. 
The probability that a node with fitness $x$ increases its out-degree by one during the step of the addition of a link  is defined:
\begin{align}\label{eq:rate_edg_discr}
\nu(x) = \frac{\sum_{l}^{N}f(x,x_l)}{\sum_{i=1}^{N}\sum_{j=1}^{N}f(x_i,x_j)},
\end{align}
that is the probability that a link is formed originating from a node with fitness $x$. The normalization factor takes in account all the possible links that could be selected. 
For large number of nodes $N$ , fitness can be considered as a continuous variable,  so we replace the sum used for discrete random variables with an integral over the set of possible values that is the average number of nodes with fitness $x$ as $N(x)=\rho(x)N$. 
So the discrete eq.\eqref{eq:rate_edg_discr} becomes:
\begin{align}\label{eq:rate_edg_cont}
\nu(x)= \frac{N\int_0^{\infty} f(x,y)\rho(y) dy}{N^2\int_0^{\infty} \int_0^{\infty} f(x,y)\rho (x)\rho(y) dx dy }.
\end{align}
The physical interpretation is that the  denominator is the total degree of the network, and  the  numerator represents the mean degree of node of fitness $x$, so defining the expected degree of a node with fitness $x$ as: 
   \begin{equation}\label{eq:expected_degree1}
    k(x)=N \int_{0}^{\infty} f(x,y)\rho(y)dy.
    \end{equation}
 Using the postulate that the link addition procedure can be considered as a sequential process, the fitness-conditional degree distribution can be found using a master equation approach where the time corresponds to the number of links added ($t=L$).  
 The degree distribution can be found using a detailed quantity, $p(k |x )$, the fitness conditional degree distribution.
A random approach  uses the expansion $ p (k |x )$ into its contributions, see \citet{boguna2003class}. 
Another approach is to derive it kinetically, where links deployed sequentially see \citet{bedogne2006complex,hoppe2015microscopic}, which fits with our approach in the network formation in the present paper. So, the probability $p_L (k|x)$  that a node  with fitness $x$ has out-degree $k$ in a network with $N$ nodes  and $L$ links  evolves as\footnote{it is possible to describe the system (see \citet{xu2010mutual}) via the rate equation for the average number of vertices with fitness $x$ and degree $k$ evolving as:
   \begin{align}
   \frac{\partial N_k (x,t)}{\partial t}=& \frac{\int_0^{\infty}f(x,y)\rho(y) dy}{\int_0^{\infty}\int_0^{\infty} f(x,y)\rho (x)\rho(y) dx dy} \left[ N_{k-1}(x,t) - N_k(x,t)\right],
   \end{align}
   with the initial condition $N_{k}(x,t=0)=\delta _{k,0}$ .}:
 \begin{align}\label{eq:master_eq_p}
p_{L+1} (k|x) =&\;\big [1 - \nu(x)\big]\,p_{L} (k|x) + \nu(x)\, p_{L} (k-1|x),
 \end{align}
 taking in account the initial condition $p_0(k|x)=\delta_{0,k}$.
 
 Solving the master equation via the generating function method we have: 
 \begin{equation}
p_L(k|x)={{L}\choose{k}} \nu(x)^k\,(1-\nu(x))^{L-k} \label{eq:propaga_discr}
 \end{equation}
 as demonstrated in the appendix \ref{sec:mastereq_a}.
 One can also use the continuous master equation that arrives to the same solution 
 of eq.\eqref{eq:propaga_discr} when $L$ is large and $\nu(x)$ is small :
 \begin{equation} %\label{eq_propag_poisson}
  p_L(k|x)\approx \frac{[L\cdot \, \nu(x)]^k}{k!}e^{-L\cdot \, \nu(x)}\label{eq:propaga_cont}
 \end{equation}
 which can also be obtained in an continuous approximation picture as discussed in the appendix \ref{sec:mastereq_c}.
Once we recovered the time prescription for the $p_L(k|x)$, also called the propagator, it is possible to write some main structural characteristics of the fitness network (see \citet{doronetwork}).
We focus our attention on a first order statistics as degree distribution and its average degree as measure of connectivity. A second order statistics is described by the  joint degree distribution and its average degree of nearest neighbor as measure of correlations among the nodes of the network.  

At this point it is possible to recover the statistical properties of  the network in the general case and also in cases where the equations collapses to the static case in particular regimes of time.
As first consideration we will restrict our calculations of the out-degree distribution since in a the study of network's response to shocks, a failures can only spread against the direction of edges, the quantity of interest here is the out-degree of a node. Anyway equivalent considerations can be addressed for the in-degree case.
At this point we apply these general characteristics to the present problem with the found  incremental network evolution.

\subsection{Generalized fitness degree transport distribution}
At this point we can have a comparison with the random generated fitness model, focusing the calculations on the particular cases of the linking function $f(x,y)$ and of the fitness distribution $\rho (x)$. 
One pretty general but still simple linking relation is a factorizable form of power law functions and  we choose the fitness distribution as power law distribution (Pareto-type):
\begin{align}
      & f(x_i,x_j)=  \frac{x_i^{\alpha}x_j^{\beta}}{A^{\alpha+\beta}}, \label{eq:linking_ab}\\
      & \rho(x)= C\,x^{-\mu}, \label{eq:fitness_distr}
\end{align}
with $x_{max}=A$ and where $C$ is the normalization factor. 
This specific choice of the fitness comes from empirical properties of real interbank networks where we observed a power law behavior in degree and size distributions, expressed in terms of sizes of bank as interbank assets for long tails behavior.  In principle we take any possible value for $\mu > 1$, differently from Montagna et Lux approach in \citet{montagna2013hubs} where $\mu =2$.
In the incremental network case, with the particular choices of $f(x,y)$ and $\rho(x)$, it is straightforward to write the relation:
\begin{align}
k(x)&=N\int f(x,y)\rho(y) dy\\
&= k_0\,x^{\alpha}
\end{align}
for the out-degree case. The constant factor $k_0$ can be calculated according to precise shape of fitness distribution (see the appendix \ref{sec:static_model}).
Since from eq.\eqref{eq_kr} $\nu(x)\propto k(x) $, we can write:
\begin{equation}
\nu(x)=r_0\,x^{\alpha}
\end{equation}

As first order statics for a given fitness $x$, we recover the most used observable in the network science that is the degree distribution.
The out-degree distribution can be calculated  from through:
\begin{equation}\label{eq:degree_kinetic}
P_L(k)= \int_{0}^{\infty} p_L(k|x)\rho(x)dx.
\end{equation}

 A novel interpretation to the network dynamics can be given also  in terms of one-dimension transport advection\footnote{Advection is usually synonymous of  linear convection} equation of fitness-packets over time giving arise to a evolving degree distribution. Let us denote as $P_L(k)=P(k,t)$ the degree distribution after $L$ links have been created (after a time $t$) and let us call $p_x(k,t)= p_L(k|x)$   the propagator representing  the conditional probability that a node  with  fitness $x$ ends up connected to $k$ nodes (see \citet{serrano2007large}).
 
The degree distribution,  in a continuous picture\footnote{the asymptotic continuous scheme treats both time $t$ and the links $k$ as continuous variables which can be considered approximately true at sufficiently long times and large networks with a pretty large number of links.}, can be seen as a comb of  fitness-packets traveling on time according  to the equations :
 \begin{subequations}\label{eq_transporteq}
 \begin{empheq}[left={\empheqlbrace\,}]{align}
  \frac{\partial }{\partial t} p_x(k,t) & = \;  -  \; \nu_x\frac{\partial }{\partial k} p_x(k,t)  \\
  P(k,t) & = \int p_x(k,t)\rho(x) \,dx
  \end{empheq}
   \end{subequations}  
  as fully analytically treated in appendix \ref{sec:mastereq_d}, where the solution is given according to the fitness packets $p_x(k,t)= \delta (k-\nu_x\,t)$.

Using the continuous transport equation it is possible to write the typcal wave evolution for any fitness distribution with a factorizable linking function (as defined in appendix \ref{sec:scalefreebehavior5}):
 \begin{align}
 P(k,t)=& \, \int \delta (k-\nu_x t) \rho (x)dx\\
 = & \; \frac{1}{A_k}\; \int \delta \Big(x- x_0(k) \Big) \rho (x) dx\\
 = & \;\frac{\rho\big(x_0(k)\big)}{A_k}
 \end{align} 
 where $x_0(k)$ is the expression of the fitness in terms of $k$ calculated from the zero of the equation $g(x)=k-\nu_x\,t=0$, and the factor $A_k= |\frac{\partial g(x)}{\partial x}|_{x=x_0(k)}$.  
  This formula provides a useful prescription to find any degree distribution after knowing which kind of the fitness distribution we use.
Choosing a factorizable linnking function, as  in the present paper, we can write $x_0(k)= k^{\frac{1}{\alpha}} / (r_0^{1/\alpha}t^{1/\alpha})$, and $A_k=\alpha r_0^{1/\alpha}t^{1/\alpha}k^{1-1/\alpha}$.

So, in the case of truncated  Pareto fitness distribution:
 \begin{align}
 P(k,t)=\;  &  \frac{C}{\alpha r_0^{1/\alpha} t^{\frac{1}{\alpha}} k^{\frac{\alpha -1 }{\alpha}}}    \Bigg(  \frac{k^{\frac{1}{\alpha}}}{r_0^{\frac{1}{\alpha}} \,t^{\frac{1}{\alpha}}}  \Bigg)^{-\mu} \nonumber \\ 
 = \;& \frac{C}{\alpha r_0^{\frac{1-\mu}{\alpha}}}\; t^{\frac{\mu-1}{\alpha}}\, k^{-\frac{\alpha +\mu-1}{\alpha}}
 \end{align} 
 the degree distribution is also  power law, as proved  in appendix \ref{sec:mastereq_d}.
 
As another example of power law degree distribution, one can choose $\rho(x)=e^{-x}$, exponential fitness, and the linking function as $f(x,y)=\Theta(x+y-C)$ where $\Theta$ is the Heaviside function and $C$ is a constant as threshold for a creation of a link. 
Such a choice brings to write $\nu_x=r_0e^{x-C}$ so that $x_0(k)=\log\frac{ke^C}{r_0t}$ and $A(k)=k$.
So, we have that the degree distribution evolution is:
\begin{align}
P(k,t)=\tfrac{r_0}{e^{C}}\,t\,k^{-2}.
\end{align}

Another case of a heavy- tailed degree distribution (see \citet{foss2013heavy}) can be obtained  considering a log-Pareto fitness distribution, so we have:
\begin{align}
P(k,t)=\;  & \frac{1}{\alpha r_0^{\frac{1}{\alpha}} t^{\frac{1}{\alpha}} k^{\frac{\alpha -1 }{\alpha}}}   C e^{-(\mu-1)  k^{\frac{1}{\alpha}} / (r_0^{\frac{1}{\alpha}}t^{\frac{1}{\alpha}}) } \nonumber \\
= \;& \tfrac{C}{\alpha r_0^{1/\alpha}}\; t^{-1/\alpha}\;   \;   e^{-c_{\alpha}\,k^{\frac{1}{\alpha}}}     \;   \; \, k^{\frac{1}{\alpha} -1}.
\end{align} 
Such distribution is a  Weibull distributions which shows heavy tails for $\alpha>1$ where the distribution is  subexponential. 
For the in-degree distribution it is sufficient to switch the parameter $\alpha$ with the parameter $\beta$.

In our model the procedure does not allow double links so we have a limit on the maximum number of links that can  be created  with a given fixed number of nodes: $L_{max}=N(N-1)/2$, that is the full connected graph (all-to-all condition).
However, in the regime of $N\to \infty$ we are always far from reaching this saturation condition,but in concrete cases with finite number of nodes, this condition can be satisfied only when $L\ll N^2$ or equivalently $\langle k\rangle \ll N$, that is the condition for which the network is called sparse.
At this point,  the solution of the degree distribution can be recover with three equivalent approaches according to a scheme where $k-$discrete and $t-$discrete as in appendix \ref{sec:mastereq_a} or $k-$discrete and $t-$continuous as in appendix \ref{sec:mastereq_c} and ultimately both $k$ and $t$ as continuous variables as in the case of eqs \eqref{eq_transporteq}.   All the analytical pictures provide the same asymptotic regime for the degree distribution as:
\begin{align}\label{asymptscalefree}
P_L (k) &\propto L^{\frac{\mu -1}{\alpha}}\; \left(\frac{1}{k}\right)^{\frac{\mu-1}{\alpha}}\left(\frac{1}{k}+O\left(\frac{1}{k}\right)^2\right) \nonumber \\
&\sim \, L^{\frac{\mu -1}{\alpha}}\, \cdot \, k^{-\frac{\alpha+\mu -1}{\alpha}},
\end{align}
which has been obtained using the different  methods as proved in  in appendix \ref{sec:kineticapprox}  and in appendix \ref{sec:mastereq_d}.

In conclusion, as regard the out-degree distribution and the indegree distribution, we recover the same scale-free behavior in the random generated model  for power law distribution of  $\rho(x)$.
\begin{align}
P(k_{out})\propto & \; k_{out}^{-\frac{\alpha+\mu -1 }{\alpha}}\\
P(k_{in})\propto & \; k_{in}^{-\frac{\beta+\mu -1 }{\beta}}
\end{align}     

Similar behaviors can be obtained for different fitness function  as truncated Pareto (also with a cutoff)  and exponential distribution as calculated in  appendix \ref{sec:static_model}.
Approaching to a dense network the power law behavior tends to disappears and the degree-distribution becomes a Gaussian around its mean value affecting also the estimation of the power law coefficient. 
In conclusion,  a scale-free behavior is present in the power law region in the degree distribution in the sparsity case, This happens when the incremental network  is run for a time  $t\ll N^2$ and so for values of degrees $1\ll k \ll N$ as shown in appendix \ref{sec_bounded}.
At this point it is possible to distinguish different regimes of evolution of such network as regard to the exponent of the linking function related to  the exponent of the fitness distribution.

A quantity of interest related are the conditional expectation of the out-degree of a randomly chosen node:
\begin{equation}\label{eq_kr}
\mathbb{E}[k|x]= k(x)=\sum_{k\geq 0}k\,p_L(k|x) = L \, \nu(x)
\end{equation}
taking in account the definition of $\nu(x)$, the unconditional average degree is:
\begin{align}
\langle k \rangle = &\int_{0}^{\infty} \mathbb{E}[k|x]\,\rho(x)dx \\
=& \;L \int_{0}^{\infty} \nu(x)\rho(x)dx ,
\end{align}
and by definition of average degree is $ \langle k \rangle =  \; \frac{L}{N}$, for $t=L\gg 1$. So we have the condition:
\begin{equation}
N \int_{0}^{\infty} \nu(x)\rho(x)dx=1.
\end{equation}

\subsection{Multigraph directed networks }

We have seen that in the incremental prescription of the fitness model if one does not set the linking probability to zero once a link between a pair of nodes has been chosen, the network is sparse, in the sense that $L\ll N^2$ , thus the possibility of adding a link to a pair of nodes that is already connected is negligible in the thermodynamic limit.
Essentially there are two different approaches on the generation of the incremental fitness evolution in the way the link are deployed among nodes: a simple graph model and a multilink version.
In the first procedure, for a given directed link, once a pair of nodes has been selected their linking  probability is set to zero, that is we do not allow other connection among two nodes  if it is already present. We have what is defined as simple graph. In the simple graph network, the master equation is valid only in the sparse condition ($N \to \infty$ or $L\ll N^2$). So, for longer  times the adjacency matrix violates this assumption and the connections start to not obey to the analytical description. Anyway it is still possible to detect natural cutoff values among which the description still describe the evolution under certain restrictions. Consequently, we encounter congestion humps in the tails of degree distributions and a change in the correlation behavior in the degree-degree probabilities.
The second alternative procedure,  we allows more than one link among pairs of nodes.  using of the full evolution described from the master equation. In this case, the  links between a pair of nodes can be considered as the strength of the connection, i.e. if a pair of nodes has three links , we can count them as a unique link with strength three.  
From this approach it is possible to generate a incremental fitness network that is directed and weighted that evolves over time. We can generate the weighted matrix from the strength of links during the evolution in different way. The simplest way is that the strength is the weight that is to  associate the number of links between a pair of nodes to the weight of their connection.
The key point is  the validity of the thermodynamic limit that essentially for finite networks coincides with the sparse condition with forbidden self-loops. 

\begin{figure}[!ht]
	\centering
	\begin{subfigure}[c]{0.6\textwidth}
		\includegraphics[width=\linewidth]{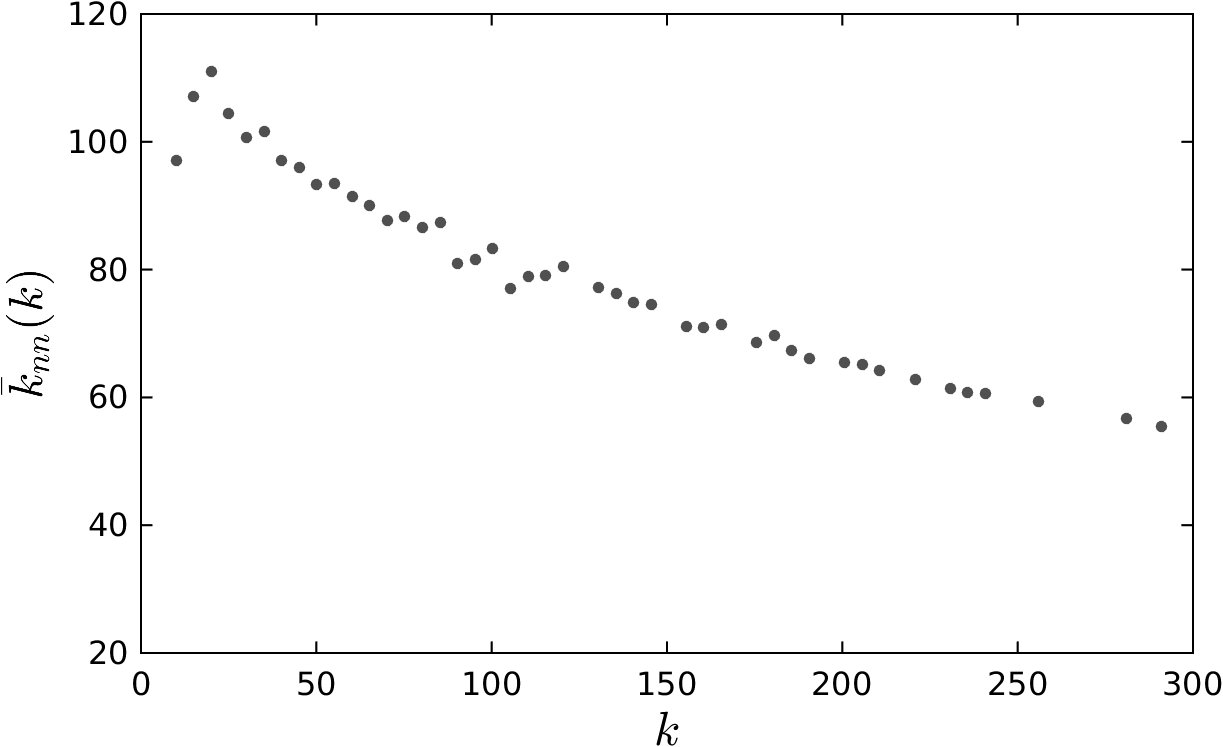}
		\caption{Assortative mixing property for simple graph network. Disassortative pattern in the out-degree connections. }
		\label{fig:avgd}
	\end{subfigure}%
	\vspace{0.05\textwidth}
	%  \qquad%\vspace{0.25\textwidth}
	\begin{subfigure}[c]{0.6\textwidth}
		\includegraphics[width=\linewidth]{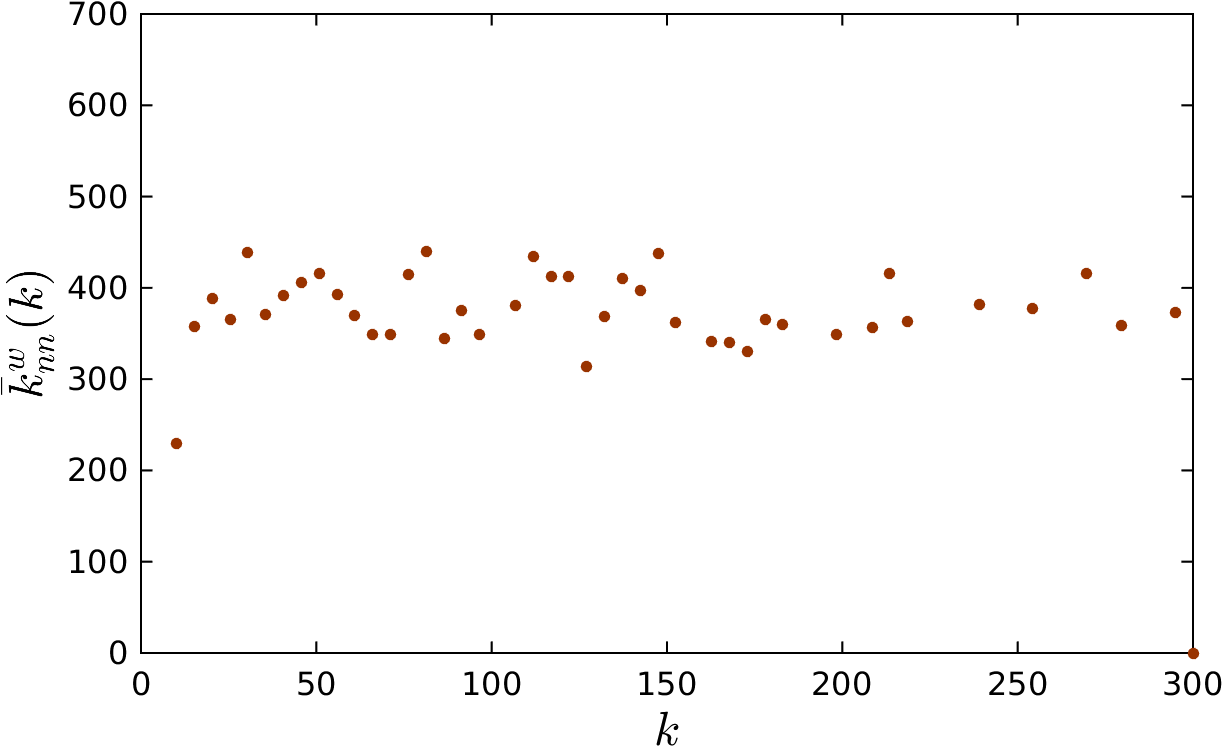}
		\caption{Assortative mixing property for multi-link weighted network. Neutral pattern for the weighted out-degree. }
		\label{fig:wavgd}
	\end{subfigure}%
	\caption{Average degree of nearest neighbors after a time of $45000$ steps. It behavior is an index of degree-degree correlation through the average degree of the nearest neighbors, $k_nn(k)$  for nodes of (out)degree $k$. In (a) we have a clear disassortative behavior for the simple graph procedure, and in (b) a neutral pattern as for the procedure with multi-links and the weighted version of the network.} \label{fig:knnplots}
	%add desired spacing between images, e. g. ~, \quad, \qquad etc. (or a blank line to force the subfigure onto a new line)
\end{figure}

\begin{figure}[!ht]
	\centering
	\includegraphics[width=0.65\textwidth]{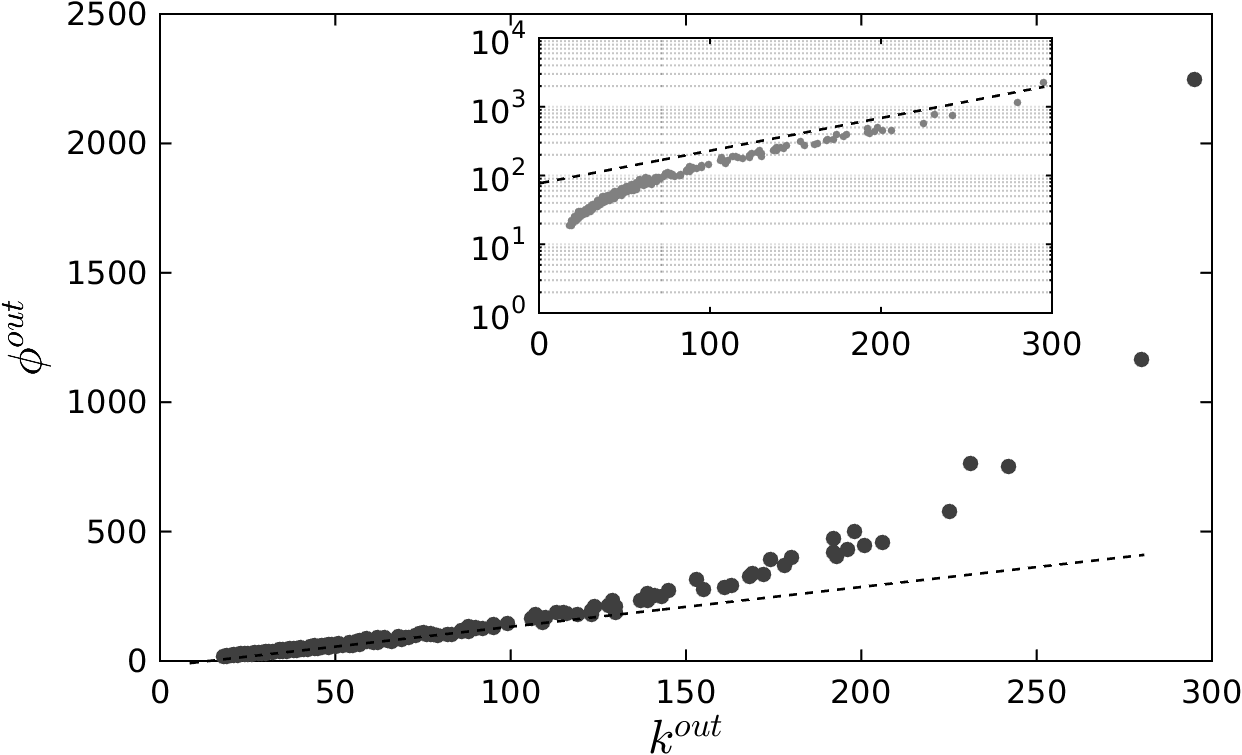}
	\caption{Strength- degree correlation in the multi-links version of the incremental fitness model. The strength of the nodes violates the uncorrelated condition of a linear dependence as clearly seen in the inbox insert in a semilog plot.}
	\label{fig:strengthedegreecorr}         
\end{figure}

In the multi-link approach  the connectivity is given by the adjacency matrix $A$ (computing the degree of nodes), and the weight matrix $W$ . The  in and out strength $\phi _i$ of a node, adding the weights of
its links is:
\begin{equation}
\phi _i^{in}(t) = \sum_j W_{i,j}(t)\quad \text{and}\quad \phi _i^{out}(t) =  \sum_j W_{j,i}(t)
\end{equation}
and similarly to the degree, the average in- and out-strength must be equal to:
\begin{equation}
\langle \phi _i^{in}(t) \rangle  = \frac{\Omega (t)}{N}=  \langle \phi _i^{out}(t) \rangle 
\end{equation}
where $\Omega$ denotes the strength of the entire network. 
In this generative model, the elements of $A$ that are already $1$ are replaced by another $1$ if their are connected again with a new connection, but the strength of the link become $2$ so changing the strength. So essentially the adjacency matrix looks the same to that from the simple graph procedure, but the structure of the network is  different because of the different strength of the network. Most important, the strength does not suffer from finite size effects as cutoff humps or change in correlation properties. 
In the case we consider the multi-link weighted generation of the incremental  fitness model, we can plot both the degree distribution and the strength distribution. As weights we directly used the number of links that  a node accumulates  over time maintaining the same linking probability toward all the nodes.
\begin{figure}[!ht]
	\centering
	\begin{subfigure}[c]{0.6\textwidth}
		\includegraphics[width=\linewidth]{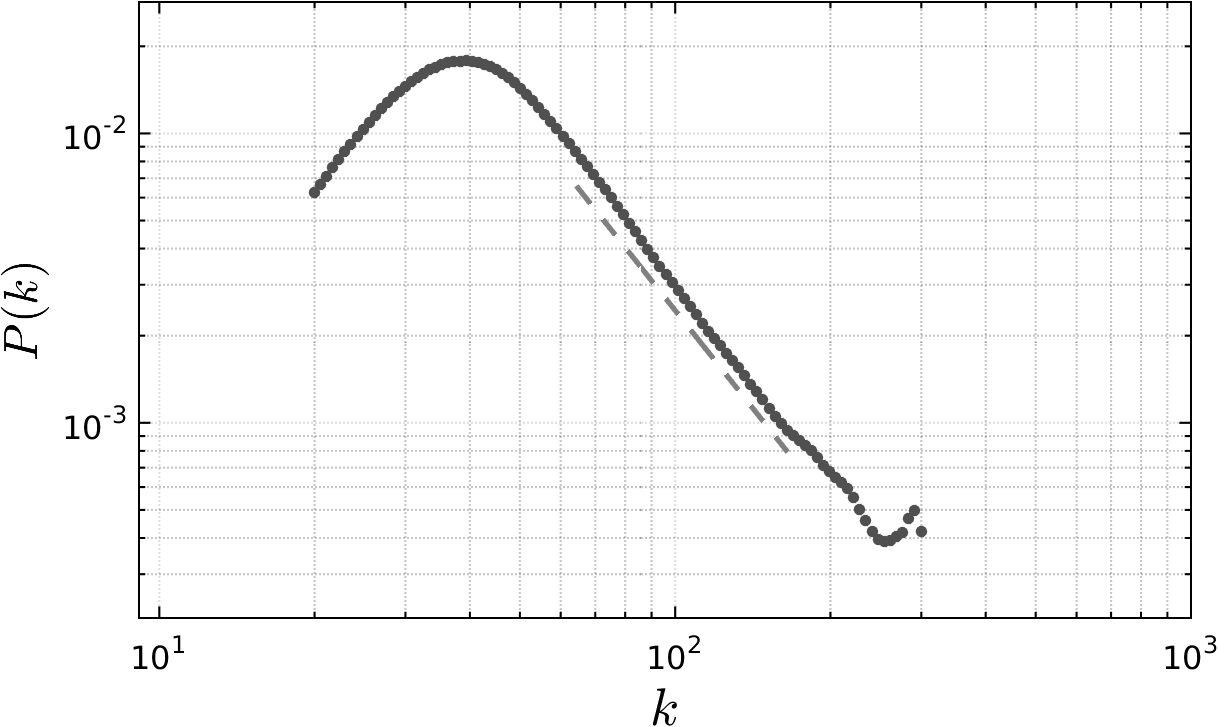}
		\caption{(Out) degree distribution. Close to a dense condition. }
		\label{fig:capacity1}
	\end{subfigure}%
	\vspace{0.05\textwidth}
	%  \qquad%\vspace{0.25\textwidth}
	\begin{subfigure}[c]{0.6\textwidth}
		\includegraphics[width=\linewidth]{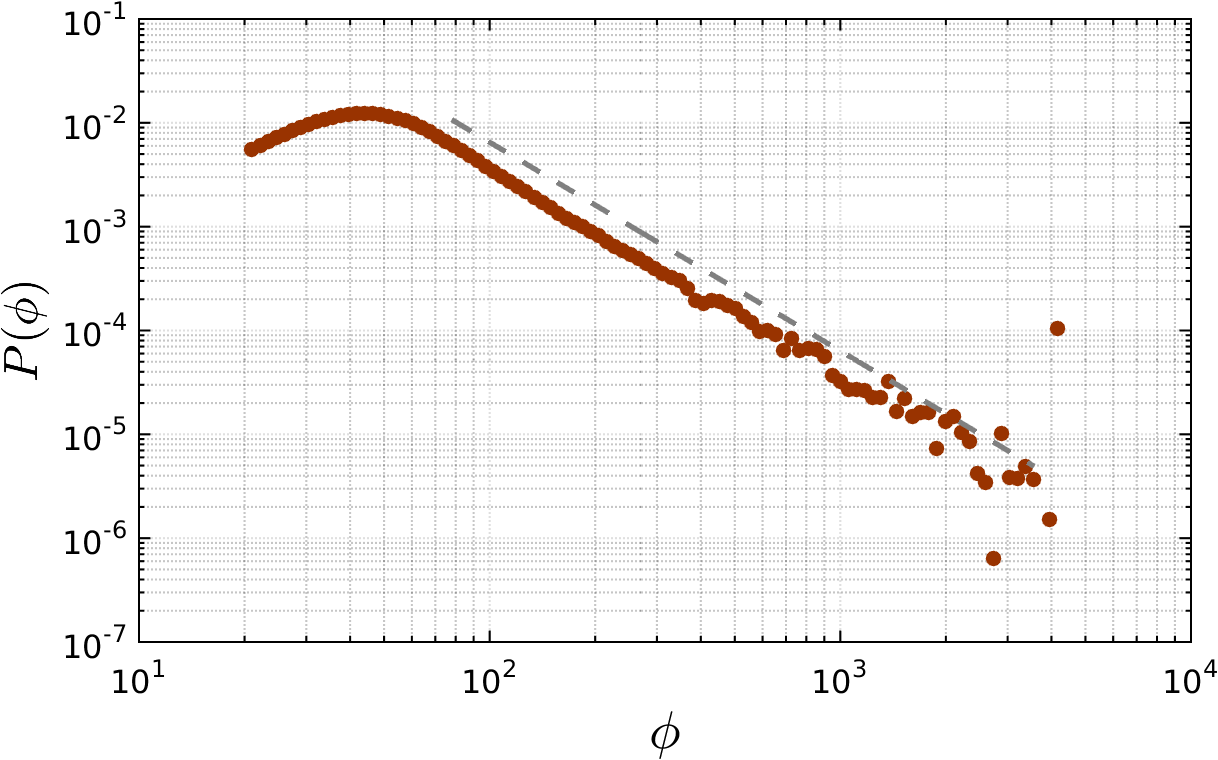}
		\caption{(Out) strength distribution. Similar  to the sparse condition. }
		\label{fig:capacity2}
	\end{subfigure}%
	\caption{Evolution of the incremental fitness model after a time of $45000$ steps. One can se that the degree distribution (a) is affected by the congestion effect due to the cutoff on the adjacency matrix. As for the strength distribution (b) there is no cutoff, since the connections continue to be added also to high degree nodes. In both figure we applied the density estimation in a logarithmic binning over 100 realizations. The dotted line in (b) represent the theoretical power law coefficients of $\tau = 2.1$ of the infinite limits, on the other side in dense networks as in (a) the slope of the power law coefficient is squeezed not revealing the exact power law coefficient due the upper bound degree. }\label{fig:capacity}
	%add desired spacing between images, e. g. ~, \quad, \qquad etc. (or a blank line to force the subfigure onto a new line)
\end{figure}
In fig.\ref{fig:capacity} we can see the two distributions where the degree distribution behaves the same as in the simple graph network, while the strength does show a clear scale free behavior over the time evolution without signs of congestion.
Moreover the basic structure of incremental simple graph network  shows  disassortativity, which is not representative of any underlying assortative or disassortative mixing. Becuase of the congestion effect, nodes with degree higher than the structural cut-off will display structural disassortativity that arises from the structure of a finite size network. 
Using multiple links  approach, the network is no longer a simple graph, but it allows for sufficient links to maintain correlation neutrality, as showed in fig.\ref{fig:knnplots}. 
If we treat the multilink network as weighted network, we can also measure the average degree of nearest neighbor degree ($\bar{k}_{nn}$) and comparing it in the version of simple graph \ref{fig:avgd} and in the weighted version for multi-link graph in \ref{fig:wavgd}.
We used $\bar{k}_{nn}$ behavior as measure of assortative structure in the network.
Defining the average nearest neighbors for weighted network is \citet{mastrandrea2014enhanced}:
\begin{align}  \label{eq:anndw}
k_{nn,i}^w= & \frac{1}{\phi_i}\sum_{j=1}^{N}A_{i,j}W_{ij}k_j
\end{align}
where $\phi_i$ is the strength of the node $i$ as already defined. Choosing the degree as out- or in- degrees we have two version of the average nearest neighbors degree.
In order to get the average degree of nearest neighbors $\bar{k}_{nn}(k)$ , we takes the average  of eq.\eqref{eq:anndw} over the number of nodes of strength $\phi$.
A way to double check the structural origin of the disassortative behavior of the simple graph model can be recover  comparing the $\bar{k}_{nn}(k)$  with a degree-preserving randomized version of itself (without multiple edges). Since there is no change in the disassortative behavior, the degree correlation pattern  is not a meaningful property of the simple graph network in the incremental fitness function. On the contrary it is an artifact in the sense that it comes from a finite size effect of the model.
Anyway, the procedure of creating a weighted network from the accumulation of links, generates nodes' strength that are not independent from the degree. This means that the weights are dependent of the network topology \citet{barrat2004architecture}, in fact im the absence of correlation the strength should grows linearly with the degree: $\phi \propto k$. In our weighted network this is violated as seen in Fig.\ref{fig:strengthedegreecorr}. The strength of nodes grows faster than their degree, denoting a correlation between the weight and the topological properties of the network. Therefore the  two quantities does not provide the same information on the system.

%On the other side, the power law coefficient is found via a maximum likelihood estimation with:
%\begin{equation}\label{eq:mel_slope}
%\hat{\tau} = 1 + n \cdot \left( \sum_{i=1}^n \ln{\frac{x_i}{x_{min}}}\right)^{-1},
%\end{equation} 
% and the lower point we have used a goodness of fit based approach via a Kolmogorov Smirnov test. 
%These estimations have been addressed with the same techniques as described in \citet{clauset2009power}. 

%As regard with the estimation of the upper point (maximum value) we used different approach for the case of truncated pareto and for exponential cutoff. As discussed in appendix \ref{appendix:estimation} for the truncated power law we used the maximum value in the real data of fitness variables, and in the degree distributions the cutoff exponential value have been evaluated via a maximum likelihood numerical estimation. The former way is straightforward and the match goods with our fitness variables of interbank assets. The former is critical because it needs to be evaluated via a numerical approach once one supposes to have already found the minimum point and the power law coefficient. It can be usually be used to find the exponential factor that is an indication of the transition between the sparse condition  the dense condition of the kinetic network model.

\subsection{Simple Finite Networks}

% % % % % % % % % % % % % % % % % % % % % % % % % % % % % % % %5
% % % % % % % % % % % % % % % % % % % % % % % % % % % % % % % % %
The incremental fitness model has a couple of  technical and practical issues to take care. One is the computational procedure to make the network  evolve in time creating new links. This determines the formation of structures in the networks that can change according to which way we consider the connections among nodes. 
The second aspect is about finite scale of the network. The choice of the coefficients of the linking function can bring to some apparent characteristic behaviors of  degree distribution, determining different scale free regimes.   
The incremental version of fitness driven networks has been described through the master equation eq.\eqref{eq:master_eq_p} that takes in account  the temporal evolution links formation in the network with  fixed number of nodes (i.e. a non growing network model). The mathematical description makes the assumption of thermodynamic limit with an infinite network. In practical case it works in the sparse condition with $L\ll N$. For longer times we encounter  the presence of multi-links that if forbidden brings to a congestion for high degree nodes and a network that become denser over time.
Recent empirical observation put emphasis on the importance of link congestion effect in networks in which density of links  increases during its evolution and  the total number of links grows faster than the number of nodes:
\begin{equation}
L(t) \propto F\left(  N(t),t\right)^a,
\end{equation}
where $N$ is number of nodes at time $t$, and $L$ the number of edges at time $t$. The coefficient $a$ is also the densification exponents. A particular choice of function $F$  in our model is for $N$  fixed over time and $a=1$, so we have $L(t)=t$.  In growing networks, also vertices are created so the function $F$ can be of the form of $L(t)\propto N(t)^a$ , and  in particular an important recent finds  \citet{leskovec2005graphs} gives the definition of densification phenomena\footnote{usually  the definition  includes both the power law behavior of number of links, both the shrinking behavior of the network diameter. } where $a>1$. In these empirical observations this phenomenon has been observed in societies and online communities as well as in the evolution of scientific fields and cities \footnote{Among the others real examples of this phenomena are:  patents citation graph, product recommendation network and large online social networks.} 
Similar kind of phenomena are also presented in terms of accelerated growth models analytically described  in \citet{PhysRevE.63.025101} with a more tractable mathematical description.
Our model is a simple and extreme model in the sense of densification effect but also focuses the attention on the effects and possible application of accelerated and densified networks.
We will focus the attention on the effects of high density  of two aspects of statistical features:  the degree distribution and the degree degree correlations.
In our case of evolution of out-degree distribution has the asymptotic form in eq.\eqref{asymptscalefree}:
\begin{equation}\label{evolution_powerlaw}
P(k,t)\sim c\; t^z\,k^{-\tau}
\end{equation}
where $z=(\mu-1)/\alpha$ for the out-degree probability, and $c$ is the proportional coefficient coming from the types of fitness distribution. We also exchange $L$ with $t$ since in our model $L=t$.
In order to have an estimation of the natural restrictions, one can impose two conditions to have the low and high degree region of the power law \citet[Ch.~14]{Bornholdt:2003} \citet{dorogovtsev2002accelerated}.
Imposing the normalization condition:
\begin{equation}\label{eq_cut1}
\int_{k_0(t)}^{N}c\,t^zk^{-\tau} \sim 1
\end{equation}
where $c$ can be calculated for a truncated pareto fitness distribution as $c\sim (1/N)^z$.
The natural initial degree $k_0$ is the value from which we have the beginning of the power law behavior of the degree distribution.

As regard as the final degree, that signs the natural cut-off  of the power law behavior.
Imposing that the probability of nodes with $k>k_{cut}(t)$ is of the order of $1/t$:
\begin{equation}\label{eq_cut2}
t\int_{k_{cut}(t)}^{N}c\,t^zk^{-\tau} \sim 1
\end{equation}

\begin{figure}[!ht]
	\centering
	\includegraphics[width=0.6\linewidth]{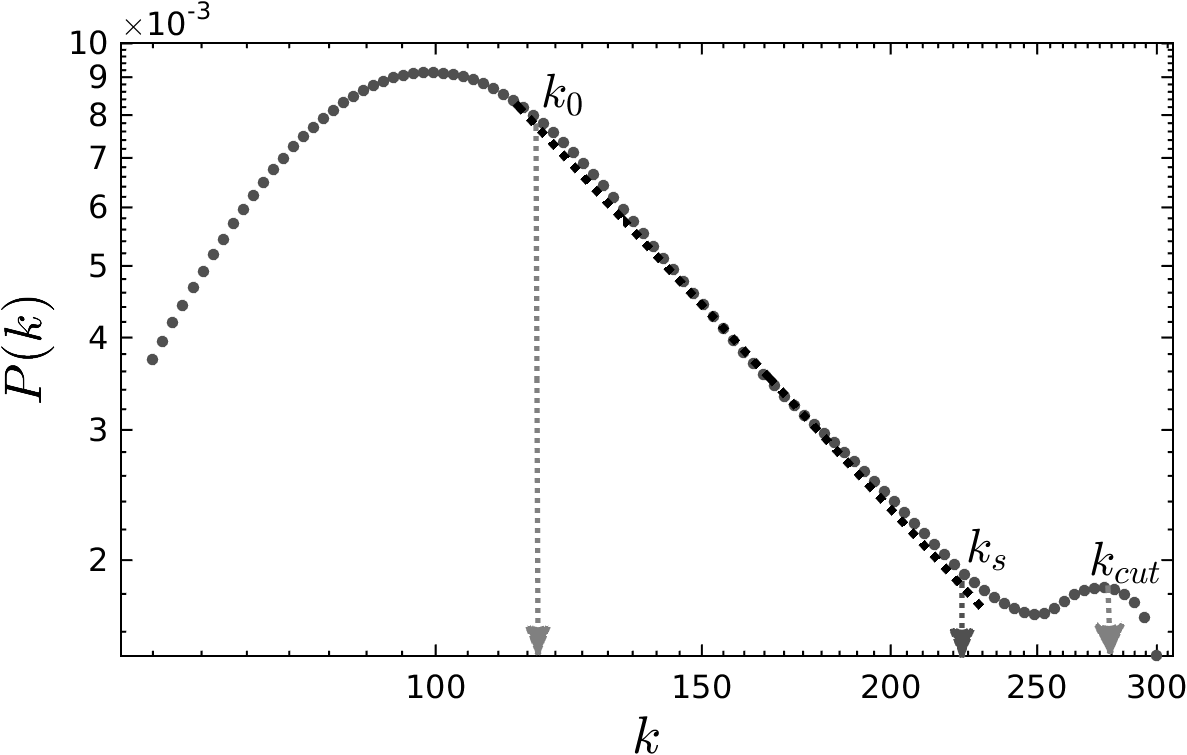}
	\caption{Probability density function of (out)degrees for the incremental fitness model, average over 100 realizations for density $\delta = 50\%$, non sparse condition.
		The black straight dotted line represents the analytical estimation of the power law region of the out degree pdf ($\tau =2.1$), and also the estimation of the natural restriction of $(k_0,k_{cut})$ as well as the structural cutoff $k_s$ in according to the analytical estimation. The evolution is made  with simple graph property. The hump near $k_{cut}$ is due to the degree congestion where high degree nodes already reached an uniform connection probability. The value for the baseline experiment are $k_0\approx 125$, $k_{cut}\approx 280$ and $k_s\approx 232$ as calculated via the eqs\eqref{eq: cutoff1} \eqref{eq: cutoff2} \eqref{eq: cutoff3}, and that corresponds what observed experimentally in the figure.       }
	\label{fig:cutoffs}
\end{figure}

The natural restriction parameters  are obtained assuming that all the elements of the sample are independently drawn from the probability density $P(k)$.
However, in real networks the degrees of the nodes are not simply independently drawn
from a probability distribution $P(k)$, but must satisfy some topological constraints due to the network structure.
It is possible to show that in order to have no correlations in the absence of multiple and self-connections, a scale-free network with degree distribution $P(k)\sim k^{-\tau}$ and size $N$ must have a cutoff scaling  called structural cutoff \citet{boguna2004cut}:
\begin{equation}\label{eq_cut3}
k_s\sim \left( \langle k \rangle N\right) ^{1/2} .
\end{equation}
Thus, if nodes of degree $k \geq k_s$ exist, it is physically impossible to attach enough links between them to maintain the neutrality of the network.
As studied by \citet{catanzaro2005generation}, this structural cutoff has the important implication that an otherwise neutral network may show disassortative degree correlations if $k_{cut} > k_s$. This disassortativity is not a result of any microscopic property of the network, but is purely due to the structural limitations of the network.
As regard with structural properties of the degree distribution, the natural cutoff has been evaluated taking in account tha the fitness distribution is a truncated Pareto-type for which we have the coefficient in eq.\eqref{evolution_powerlaw} is:
\begin{equation}
c=\frac{\rho_0 r_0^z}{\alpha}\simeq \frac{\alpha}{N^z}
\end{equation}
In the specific case of our simulation we calculate  natural cutoffs of eqs.\eqref{eq_cut1}\eqref{eq_cut2}\eqref{eq_cut3}: 
\begin{align}
k_0=&  \sqrt[{1-\tau}]{(1-\tau) \left[   \frac{(N-1)^{1-\tau}}{1-\tau}- \left( \frac{N-1}{t} \right)^z \,\right] }   \label{eq: cutoff1}\\
k_{cut}=& \sqrt[{1-\tau}]{(1-\tau)  \left[   \frac{(N-1)^{1-\tau}}{1-\tau} - \left( \frac{N-1}{t} \right)^z \cdot\frac{1}{t}  \right] }    \label{eq: cutoff2}\\
k_s=& \sqrt{\frac{L}{N}N}=\sqrt{L}  \label{eq: cutoff3}
\end{align}
In the previous equations, if we study out-degree distributions $\tau =(\alpha +\mu -1 )/\alpha$ and if we study the in-degree distribution $\tau = (\beta +\mu -1 )/\beta$. 
An example of the estimations in the baseline case is given in fig.\ref{fig:cutoffs}, in the simple graph case where multilinks are not allowed.
%It is interesting to notice that both $k_0$ and $k_{cut}$ are degree cutoff in common to the static and the kinetic version of the model. 
%On the contrary, the structural cutoff $k_s$ is central for the kinetic procedure since we can move the time forward up to the desired number of links. For the static version, the number of links has a mean value connected to the average degree so $k_s= \sqrt{\langle k\rangle N}$.

\section{Properties of incremental model}
In the entire discussion we have dealt with power law functions both in the probability function of fitness variables and in the degree distributions of the generated network. 
Once the model have been set, we want to show the usefulness of it showing how real world systems can be represented through this network. A delicate issue is the estimation of these network parameters via fitting procedures in order to generate simulations that mimics data observed in the real world as in financial markets (\citet{BODEVA2016}). 
%As real data we will use financial information of banks regarding their consolidation level. Those data on fitness variables builds on the data provided by Bankscope database as discussed in \citet{BODEVA2016}.
\begin{figure}[!ht]
	\centering
	\includegraphics[width=0.75\linewidth]{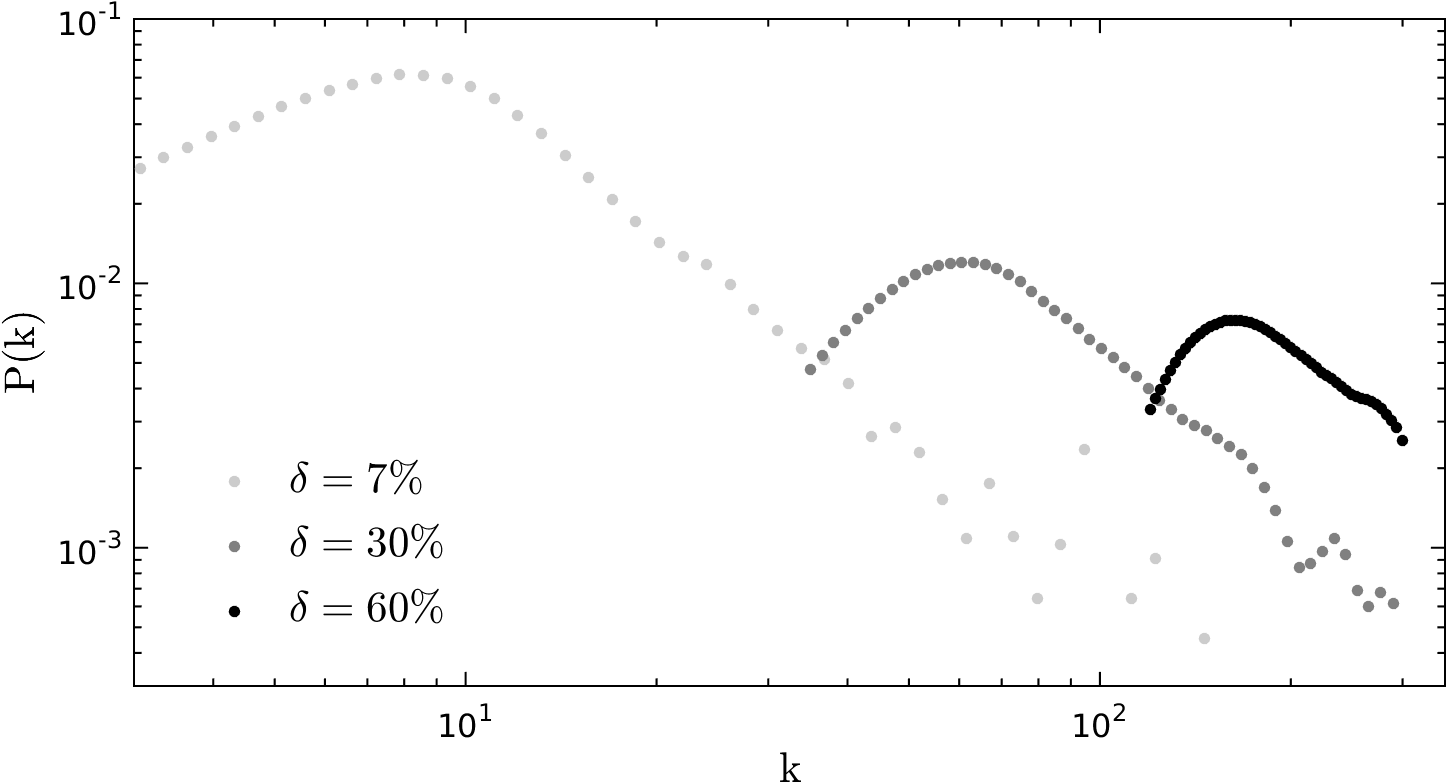}
	\caption{Probability density function of (out)degrees for the incremental fitness model for a single trajectory. The evolution excludes multi-links. Pdfs have been calculated using a logarithmic binning and a kernel density estimation with $100$ Montecarlo replications. Each function corresponds to a different density $\delta$ of the network. Increasing levels of grays corresponds to an increasing number of links where the evolution of the network last for a longer time. The asymptotic behavior is a power law distribution in the sparse condition. As the time passes and more links are added, a condensation peak appears for high degree nodes that cannot increase more their degree  because of the finite size of the network.}
	\label{fig:sparsedense}
\end{figure}
In our simulations we used a baseline setup the model as in Table \ref{baseline_table} where the  fitness function is a truncated Pareto power law and the linking probability  is a factorizable function. 
% \usepackage{booktabs}
\begin{table}[]
	\centering
	\caption{Baseline Model, the number of links changes according to the chosen average degree with $L=\langle k \rangle N$.}
	\label{baseline_table}
	\begin{tabular}{@{}lcr@{}}
		\toprule
		Parameter & Value & description \\ \midrule \midrule
		$N$       & $300$ & \textit{number of nodes} \\ \midrule \midrule
		$a$       & $5$  & \textit{fitness Pareto  lower bound}  \\ \midrule
		$b$       & $299$ & \textit{fitness Pareto  upper bound} \\ \midrule 
		$\mu$     & $1.99$ &\textit{ fitness Pareto  coefficient}\\ \midrule \midrule
		$\alpha$  & $0.7$ & \textit{linking outdegree coefficient }\\ \midrule
		$\beta$   & $0.9$  & \textit{linking indegree coefficient} \\  \midrule
		$A$   & $299$ & \textit{maximum fitness}\\ \midrule \midrule
		$\tau_{out}$ & $2.41$ &  \textit{outdegree distribution coefficient} \\ \midrule
		$\tau_{in}$ & $2.10$ & \textit{ indegree distribution coefficient} \\  \bottomrule
	\end{tabular}
\end{table}

\paragraph{Degree distribution}
A concept related to the average degree is the network density $\delta$, useful to control the number of generated links in comparison to all the possible ties:
\begin{equation}
d =\frac{\langle k \rangle}{N-1} = \frac{L}{N(N-1)}
\end{equation} 
well defined for finite network where the average degree is always finite. According to this index we have the sparse condition on the network if $d \ll 1$.

Moreover the in and out degrees are computed as:
\begin{equation}
k_i^{in}(t) =  \sum_j A_{i,j}(t) \quad \text{and}\quad  k _i^{out}(t) =  \sum_j A_{j,i}(t)
\end{equation}
where $A$ is the adjecency matrix. The average degrees are
\begin{equation}
\langle k _i^{in} (t)\rangle  = \frac{L(t)}{N}=  \langle k _i^{out}(t) \rangle .
\end{equation}
The incremental fitness model generates different degree distribution along the time prescriptions of adding links that corresponds to an increase of density rate as $\delta = d\times 100\%$  where $t=L$ and $N$ is fixed.
As shown if Fig.\ref{fig:sparsedense}, as time passes it is always possible to see the power law behavior of degree distribution, but the scale-free region becomes more and more restrictded while the condensation to a dense network become dominant. The dense condition is always present for high degree nodes but for strong connected network a peak shape in the tail becomes more and more evident reflecting the dense condition described in appendix \ref{sec:kineticapprox}.
A translation of the offset is due to the $L$ coefficient in eq.\eqref{asymptscalefree}, and low degree nodes becomes more and more rare while  links are added.
The slope of such distribution has been evaluated according to the maximum likelihood estimation.
We proved the equivalence of the degree distribution among the incremental prescription proposed here and the random prescription as in \citet{caldarelli2016data}. For a fitness distribution with a  pareto-like distribution we found that evolutionary degree distribution eq. \eqref{asymptscalefree} coincides with the random version (calculated in appendix \ref{sec:static_model}, in particular eq.\eqref{eq_truncatedDegreeProb}).  Throughout the paper we have always used the truncated Pareto distribution as a fitness distribution togheter with a factorizable linking function. Anyway similar results in terms of scale-free degree distribution   are obtained if we use fitness values taken from a  power law distribution with exponential cutoff and from exponential distribution see appendix \ref{sec:scalefreebehavior2} and \ref{sec:scalefreebehavior3}. Anyway one should also take in account different regime behaviors of power law distributions  in evaluating the fit of such  distributions as discussed in the appendix \ref{sec_regimes}.

\begin{figure}[!ht]
 \centering
                  \includegraphics[width=0.7\linewidth]{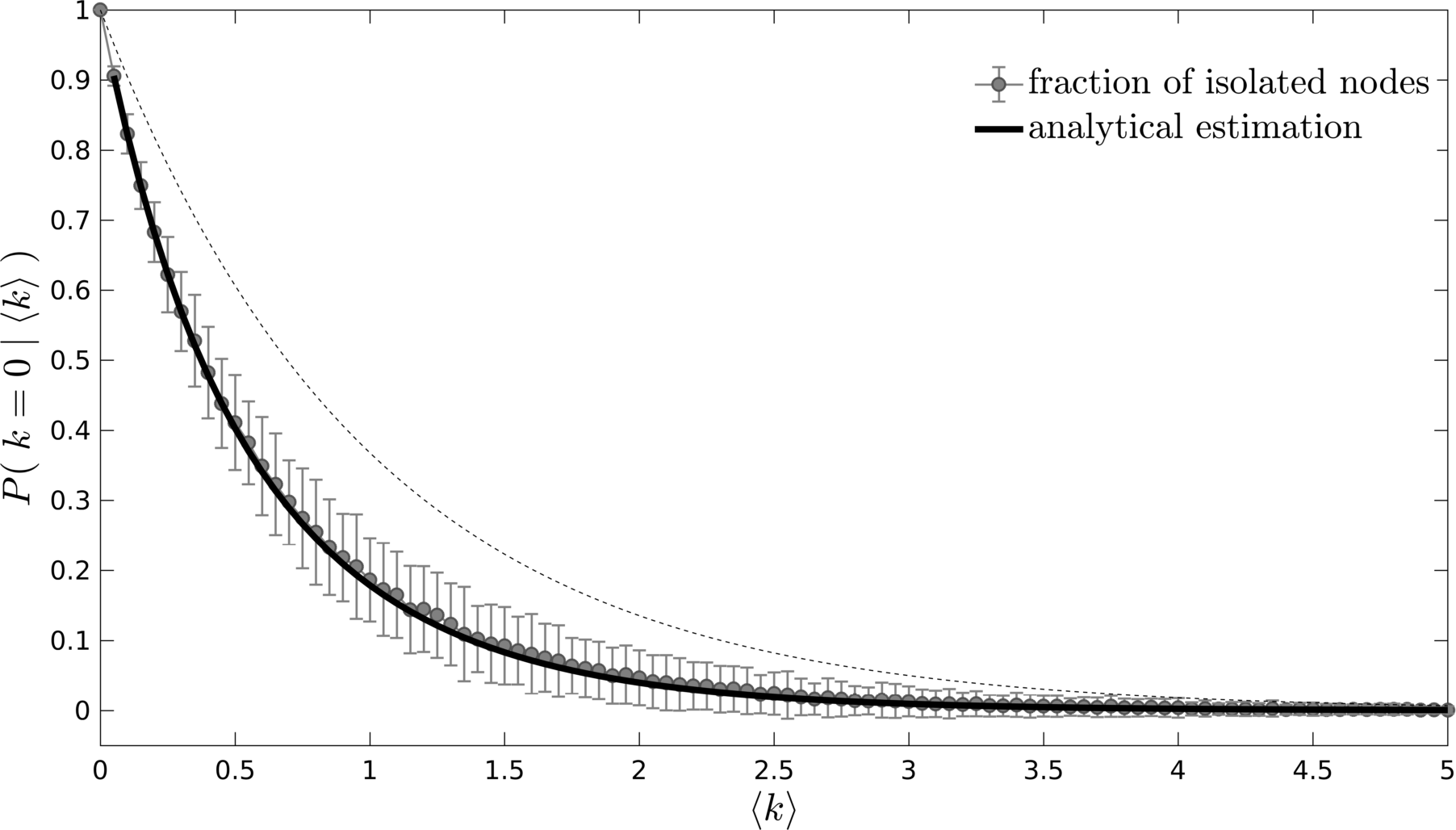}
                  \caption{ Probability to find that a randomly chosen node is isolated in a network with a given average degree. This is exactly the fraction of empty nodes over the total number of nodes that is computed with a Montecarlo simulation with $100$ repetitions, which is represented in the figure with a barplot, where the bar size is two standard deviations large. The black solid line is the analytical estimation of such fraction and the dotted gray line is the same fraction in the random graph estimation with exponential decay. }
                  \label{fig_emptynodes}
 \end{figure}

\paragraph{Isolated nodes}
For practical purpose, it is important to estimate the number of nodes with zero links, which are called isolated since in the case of network contagion or failure cascades they are not affected and do not affect the system. 
During the incremental evolution of the networks with $N$ nodes, not every node has at least a connection with others nodes. We analytically estimate the average number of isolated nodes along the time, i.e for different values of average degree that is for different numbers of created links.
In general, the probability that a node with fitness $x$ has degree $k$ after $L$ links have been created (after time $t$) is given by eq.\eqref{eq:propaga_discr}, so, since we are interested in the isolated nodes, we set $k=0$ so can write:
%$$
%p_x(k|L) = \binom{L}{k} \nu(x)^k\Big(  1-\nu(x) \Big)^{L-k}
%$$
\begin{equation}
p_x(0|L)=(1-r_0x^{\alpha})^L
\end{equation}
for the outoging degree case.
Knowing that our fitness distribution is a truncated power law distribution with  $x \in [a,b]$, we can estimate the total number of nodes with no outgoing links we can write: $$P_{out}(0|L)=\int_{a}^{b}(1-r_0x^{\alpha} )^L\; C x^{-\mu} dx.$$
At this point we can use the Chebyshev theorem on the integration of binomial differentials to solve the integral $$I_{\alpha}=\int x^{-\mu}(1+\beta x^{\alpha})^{L}\,dx$$  and  which  is expressible by of the elementary functions   if $L\in \mathbb{Z}$, which is our case since the total number of links $L$ is an integer number.
Defining $y=\beta x^{\alpha}/1$  and $\beta = -r_0$ we can write the  antiderivative as:
$$
I_{\alpha}={1\over \alpha}\beta^{-{-\mu+1\over \alpha}}B\left(y;{ 1-\mu\over \alpha },L-1\right)
$$
where $B$ is the incomplete Beta Function (\citet{chaudhry1994generalized}).
We can also  find the probability for incoming empty nodes as $I_{\beta}$ and then multiply the two probabilities if we want to find the case of fully isolated nodes completely disconnected from the network since they do not have any links at all.
%The probability that a node of fitness $x$ has no links is given by:
%$$
%p_x(k=0,t)=(1-\nu(x))^t
%$$
So the probability to find isolated nodes when there are $L$ links is given by:
\begin{align}\label{eq_isonodi}
P(0,L)= & \int_a^b p_x(0,L) \rho (x)dx\\
= \;& I_{\alpha}I_{\beta}
\end{align}
where we used
$$
I_{\alpha} =\frac{C}{\alpha} (-r_0)^{\frac{1-\mu}{\alpha}} \Bigg[   B(-r_0 x^{\alpha} ; \frac{1-\mu}{\alpha},L-1) \Bigg]_{x=a}^{x=b} 
$$
for the outoging links and similarly we define $I_{\beta}$ for the incoming links. 
Then, numerically  inverting this equation it is possible to know how many links we need to generates in order to have a certain percentage of isolated nodes in the network.
In some practical application, it can be very useful to know the expected number of nodes with no connections at time $t$, i.e. at a given average degree for the network. For example, in the case of shock propagation in network, disconnected nodes  do not contribute to the cascade, in particular if selected as first step of the cascade,  isolated nodes  cannot trigger any shock propagation. 
A rough estimation of empty nodes after having $L=\langle k \rangle N$ links can be found in the random hypothesis of the probability of the allocation of undistinguished links in $N$ nodes $(1-1/N)^{L}$ so, the excepted fraction of isolated nodes is of the order of  $e^{-\langle k \rangle}$. 
We can evaluate the expected number of isolated nodes, running a Montecarlo simulations results over $100$ realizations and compare the results with the analytical evaluation of isolated nodes after $L$ links   with the eq.\eqref{eq_isonodi}.  We see, in Fig.\ref{fig_emptynodes} how the analytical calculations predict the probability to find a given number of empty nodes for different values of the average degree.    In contrast, the  random hypothesis estimation gives very loose results at least for intermediate values of average degrees. 

\begin{figure}[!ht]
	\centering
	\includegraphics[width=0.6\linewidth]{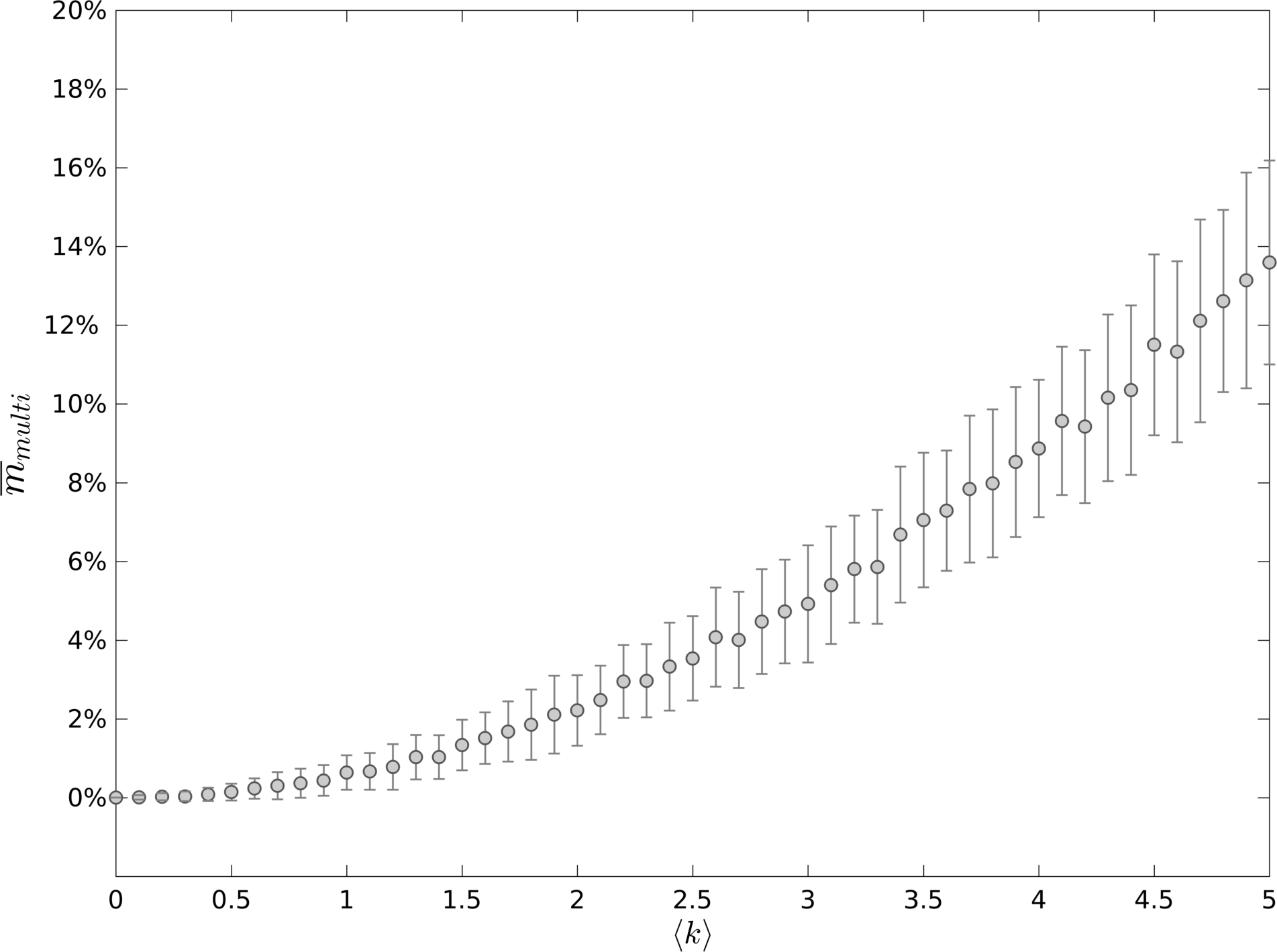}
	\caption{ Average fraction of multilinks in the network for different values of the average degree and so number of links in the system.  The average is obtained with $100$ replications. }
	\label{fig_multilink}
\end{figure}

\paragraph{Expected number of multiple links}
As a measure of difference between the multigraph and simple graph procedures, it could be useful to estimate of the expected number of multiple links in the graph. Following the combinatorial argument used by \cite{latora2017complex}, the expected number of multiple links in the network is:
\begin{equation}
\overline{m}_{multi}= \frac{1}{4}\left( \frac{ \langle k^2 \rangle -\langle k \rangle  }{\langle k \rangle} \right)^2
\end{equation}
which can be verified both for out and for in degree distributions. In our baseline model the first moment of the distribution is finite, on the contrary the second moment diverges (see appendix \ref{sec_regimes}) so that the number of multilinks tends to  increase with the number of links as seen in Fig.\ref{fig_multilink} with  Montecarlo simulations.

\begin{figure}[!ht]
	% \centering
	\begin{subfigure}[t]{.4\textwidth}
		\includegraphics[width=\linewidth]{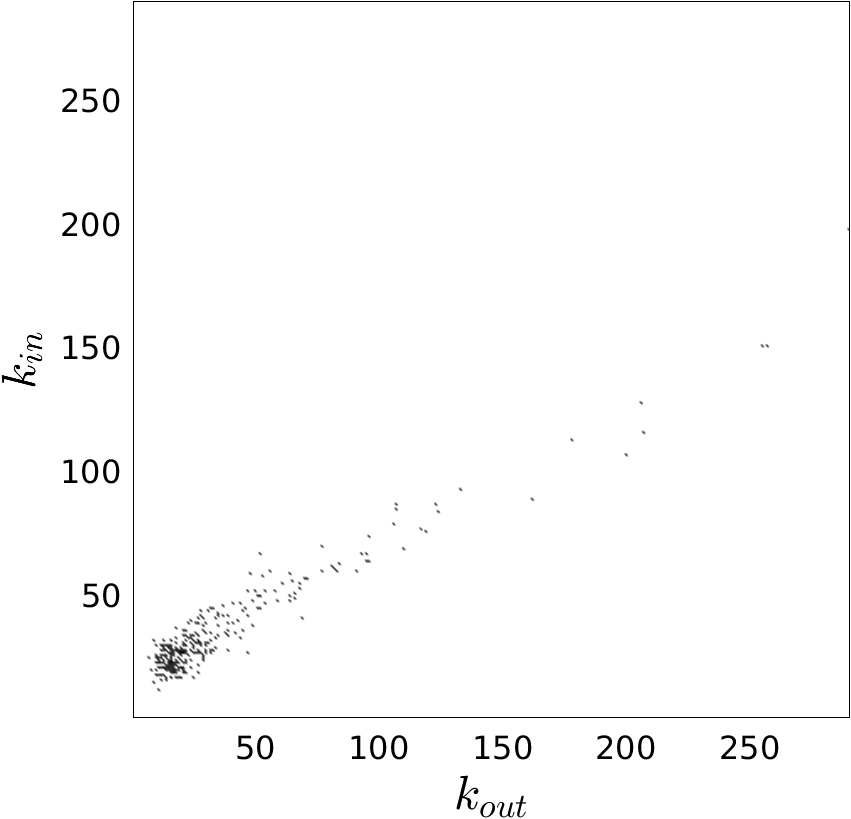}
		\caption{The number of nodes in the network with given in-degree and out-degree $\mathbf{k}=(k_{out},k_{in})$. This is proportional to the  joint degree distribution $P(k_{out},k_{in})$  which probability of a randomly picked node  that has in-degree $k_{in}$ and out-degree $k_{out}$ .}
		\label{fig:joint}
	\end{subfigure}%
	% \vspace{0.05\textwidth}
	\qquad%\vspace{0.25\textwidth}  	       
	%   \centering
	\begin{subfigure}[t]{.45\textwidth}
		\includegraphics[width=\linewidth]{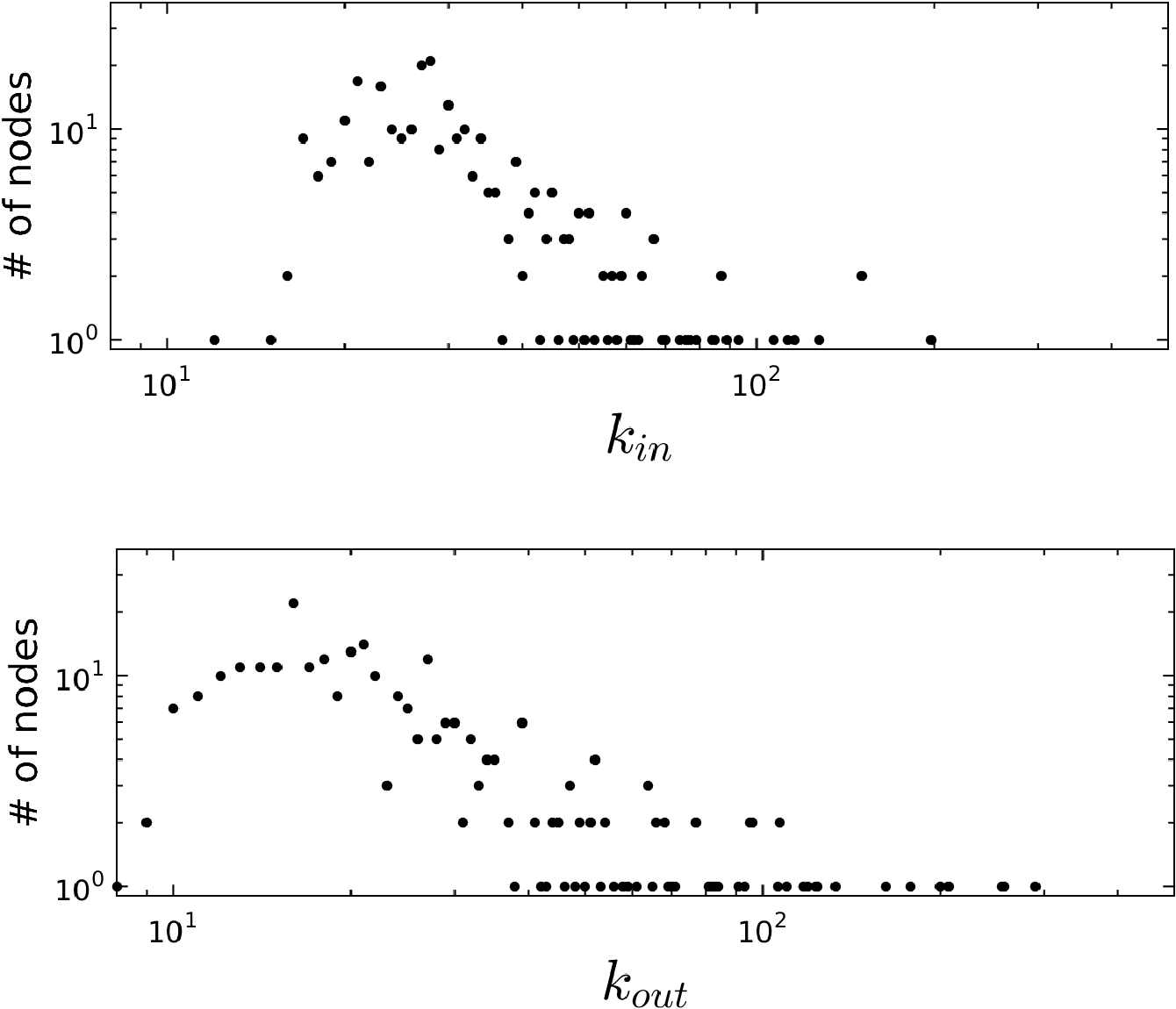}
		\caption{ number of nodes with given in/out degree. They indicate  the marginal probabilities  $P(k_{in})$ and $P(k_{out})$.}
		\label{fig:marginal_out}
	\end{subfigure}%
	% \vspace{0.05\textwidth}
	%            
	%            \medskip
	%            
	%            \centering
	%            \begin{minipage}[c]{.4\textwidth}
	%            \text{ }
	%            \end{minipage}         
	%               \centering
	%              	 \begin{subfigure}[t]{0.4\textwidth}
	%                   	                  \includegraphics[width=\linewidth]{immagini/marginal_outdegree.png}
	%                   	                  \caption{number of nodes with given out degree. It's proportional to the marginal probability $P(k_{in})$.}
	%                   	                  \label{fig:marginal_in}
	%                            \end{subfigure}%
	
	\caption{Degree distributions in the directed network. In (a) the bidimensional joint degree distribution, and (b)  represents the marginal distribution along the direction of out and in degree, as in fig.\ref{fig:sparsedense}. In this case the density of the network is $\delta=7\%$ and the binning is equal size without any density estimation.}\label{fig_bidegree}
	%add desired spacing between images, e. g. ~, \quad, \qquad etc. (or a blank line to force the subfigure onto a new line)
\end{figure}

\paragraph{Correlations and assortativity}
% % % % %   JOINT - Marginal DEGREE  % % % % % % % % % %
Higher order statistics in network involve joint degree distributions and conditional probabilities that contains degree correlation information.
In a directed graph, another form of degree correlation  is otherwise expressed by the  joint in/out-degree probability distribution function $P(\mathbf{k})\equiv P(k_{in},k_{out})$. 
Let us define the transition probability  $P_{in}(\mathbf{k}' | \mathbf{k})$ as the probability of reaching a node of degree $\mathbf{k}'$ starting from a node of degree $\mathbf{k}$ through an incoming link; and a similar definition for the  transition probability  $P_{out}(\mathbf{k}' | \mathbf{k})$ along an outgoing link. These transition
probabilities are related through the following degree detailed balance conditions  \citet{boguna2005generalized}
\begin{equation}
k_{out}P(\mathbf{k})P_{out}(\mathbf{k}' | \mathbf{k})= k'_{in}P(\mathbf{k}')P_{in}(\mathbf{k}| \mathbf{k}')
\end{equation}
These conditions assure that any link departing from a node, always points  to another node (network closeness).  We will discuss numerical behaviors of the joint degree distribution in the results $P(k_{in},k_{out})$ using Montecarlo simulations.
In the present discussion, the network is directed, the edges are differentiated into incoming and outgoing, so that each vertex has two coexisting degrees $k_{in}$ and $k_{out}$ , with total degree $k = k_{in} + k_{out}$. Hence, the degree distribution for a directed network is a joint degree distribution $P (k_{out}, k_{in}) \equiv P (\mathbf{k})$ of in- and out-degrees that in general may be correlated. 
Real networks are usually correlated in the sense that the degrees of pairs of connected nodes are correlated random quantities. When two-point correlations are absent, the transition probabilities become independent of the degree of the source node, we have:
\begin{align}
P_{out}(\mathbf{k}' | \mathbf{k})=&\;\frac{k'_{in} P(\mathbf{k}')}{\langle k_{in}\rangle} \\
P_{in}(\mathbf{k}' | \mathbf{k})=&\;\frac{k'_{out} P(\mathbf{k}')}{\langle k_{out}\rangle} 
\end{align} 
which is the case of absence of degree-degree correlations.
%to use bar plots, we could look at the marginal degree distributions.  One of the marginal degree distributions is the in-degree distribution.
The marginal degree distributions $P(k_{in})$ and $P(k_{out})$ involving just the in-degree and just the out-degree are a lot simpler to deal with. So, rather than dealing with the full two-dimensional degree distribution, one could just study the marginal distributions separately. In Fig.\ref{fig_bidegree} we apply the previous discussion to our fitness network model. We can see how the in-out joint distribution is uncorrelated and each marginal probabilities  are those find in the previous sections, in particular if in the linking function $\alpha = \beta$ the points in the Fig.\ref{fig:joint} should stay in the diagonal. 
By construction,  there are no  correlations between a node's in-degree and a node's out-degree, in the sense that if one node has a large in-degree and another nodes has a small in-degree, both nodes are equally likely to have a large out-degree. The strong influence of the correlation between in- and out-degree can be seen by the fact that it determines the largest eigenvalue of the adjacency matrix \citet{restrepo2007approximating}. In our network, the assumption of uncorrelated in- and out-degrees  make sense, but it certainly doesn't have to be true in general.

\begin{figure}[!ht]
     \centering
     	 \begin{subfigure}[c]{0.7\textwidth}
     	                  \includegraphics[width=\linewidth]{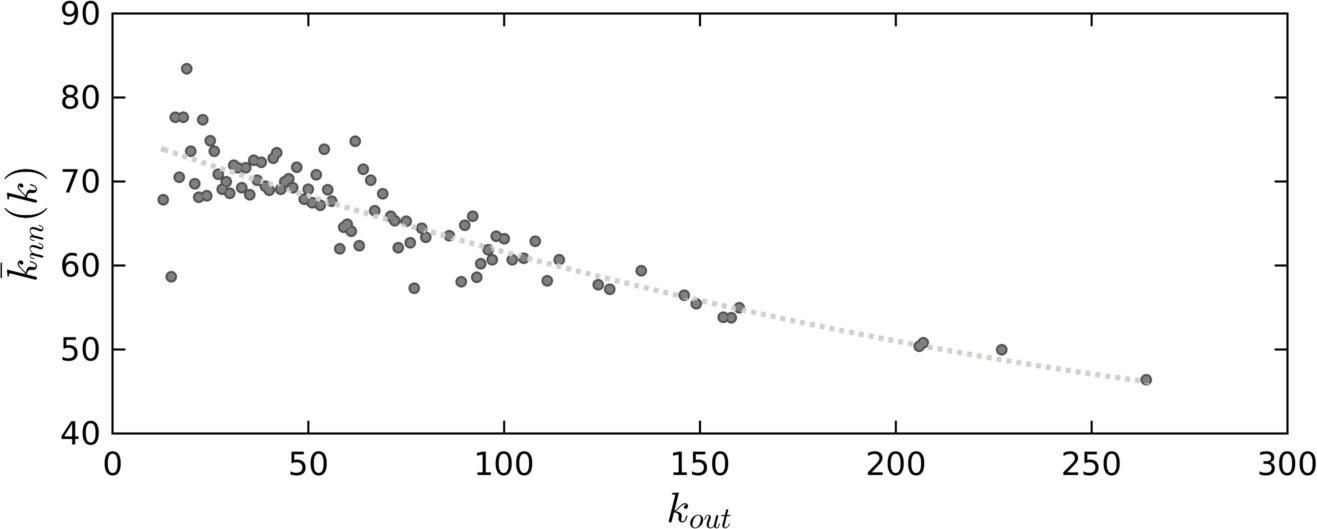}
     	                  \caption{Outgoing ANND  }
     	                  \label{fig:knn_out}
              \end{subfigure}%
                 \vspace{0.05\textwidth}
              	 \begin{subfigure}[c]{0.70\textwidth}
                   	                  \includegraphics[width=\linewidth]{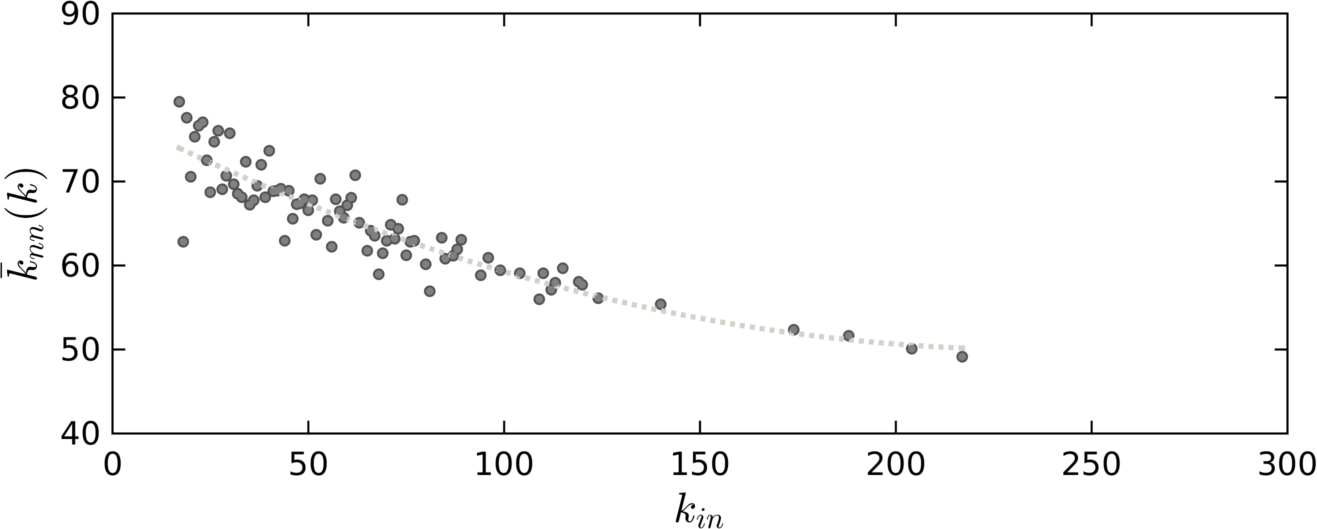}
                   	                  \caption{Incoming ANND}
                   	                  \label{fig:knn_in}
                            \end{subfigure}%
              \caption{Average degree of nearest neighbors and disassortative behavior of the incremental fitness model in our baseline but in a very dense network with 45000 links ($\delta = 50 \%$).}\label{fig:bidegree}
             %add desired spacing between images, e. g. ~, \quad, \qquad etc. (or a blank line to force the subfigure onto a new line)
     \end{figure}

\begin{figure}[!ht]
 \centering
                  \includegraphics[width=1\linewidth]{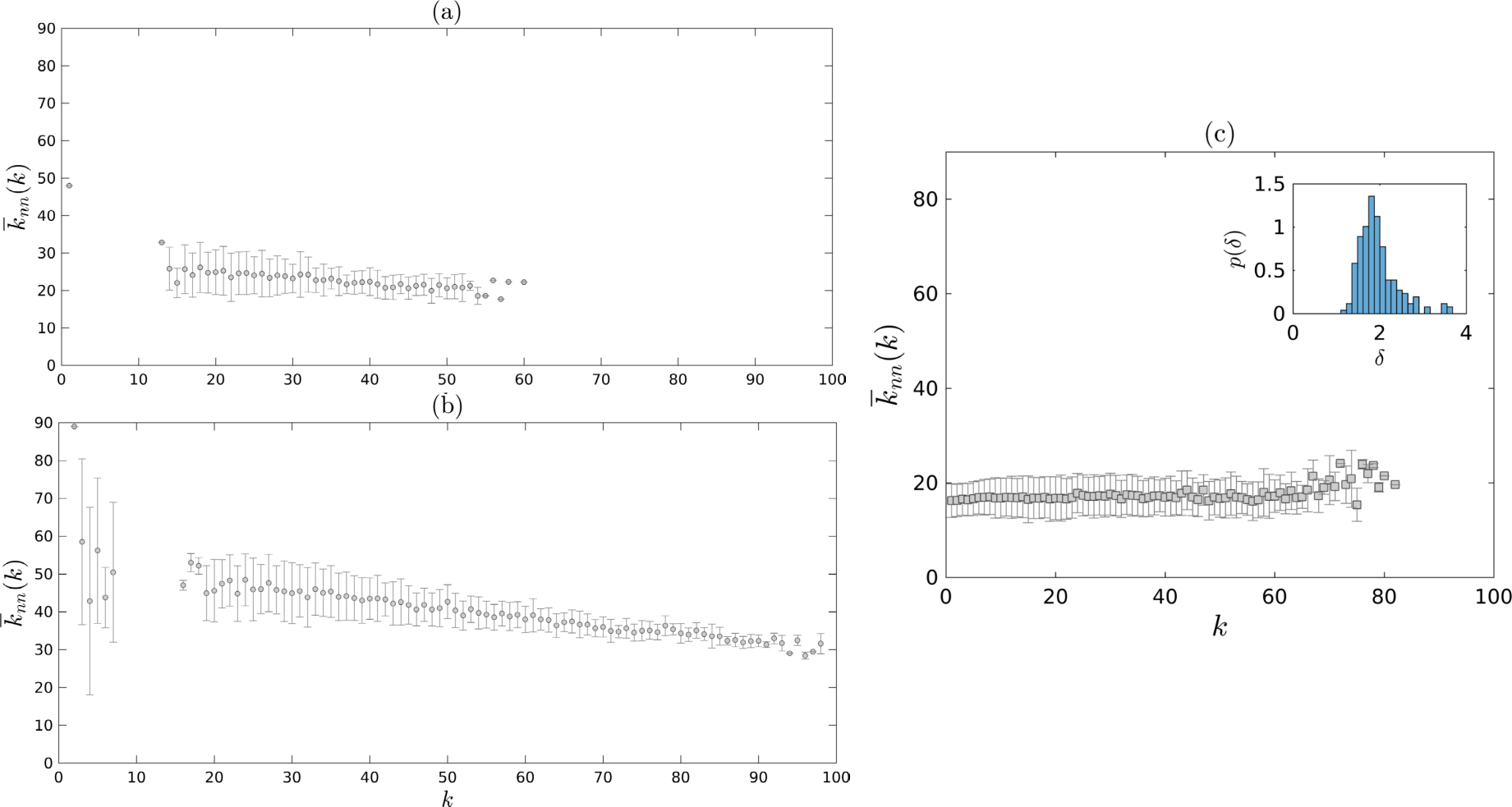}
                  \caption{ANND function in the incremental model (a) and (b)  and in the random model (c) over averaged over $200$ Montecarlo replications. In particular: (a) the network is run up to reach a density of $\delta = 10\%$ where the assortativity is almost neutral within  the the error bars. (b) the network reached a density of $\delta = 20\%$ where the disassortativy behavior is more evident. On the contrary, in the random  model the network, replicated $200$ times, shows a $\delta$ varying according the random configuration reached for each replication. So $\delta$ is distributed according to the inset in the figure and the assortative mixing results to be neutral. Actually, for single realizations, the two models show the same spurious disassortativity for finite networks. }
                  \label{fig_assortaKS}
 \end{figure}

Since the study of network's response to shock a  failures can only spread against the direction of links, % the quantity of interest here is the out-degree of a node.
%In order to consider percolation in this network, 
it is necessary to obtain the fitness-conditional out-degree distribution (the propagator). As  discussed in \citet{serrano2007correlations},  second order statistics deals with degree-degree correlations which its empirical measurement is the average neighbour connectivity $k_{nn}(k)$, measuring the average degree of nodes neighbor of nodes with degree $k$.
A seminal work on correlations in hidden-variable networks have been done by \citet{boguna2003class} where the authors found analytical expressions for the main topological properties of these models as a function of the distribution of hidden variables and the
probability of connecting nodes.
As a general discussion we consider the total degree $k=k_{in} + k_{out}$, 
%In the general case of directed network 
so, the joint degree distribution $P(k,k')$ is  the probability that a randomly chosen link connects nodes of degree $k$ and $k'$.
Moreover the conditional probability $P(k'|k)$ is the probability that if one end-node of a link is of degree $k$, then its second end-node is of degree $k'$.
The relation of those probabilities is expressed through the degree detailed balance condition:
\begin{equation}
kP(k'|k)P(k)=k'P(k|k')P(k')=\langle k \rangle P(k,k')
\end{equation}  
Actually, such histograms in finite size systems are highly affected by statistical fluctuations getting  bad   empirical evaluate of network correlations. As better characterization of degree correlations, it is then more convenient to use the mean degree of the neighbors of a node as a function of its degree $k$.
The average nearest neighbors degree (ANND), $\bar{k}_{nn} (k)$, of vertices of degree $k$ is defined as a smoothed conditional probability:
\begin{equation}
\bar{k}_{nn} (k) = \sum_{k'} k'P(k'|k)
\end{equation}
so that the  statistical fluctuations of $P(k'|k)$ results to be more dumped.
In the case of incremental fitness driven networks, we have the hidden variable $x$ that represents a property of that node influencing its ability to make links.
In our case of fitness driven network \citet{boguna2003class,serrano2007large,servedio2004vertex} let us  define the conditional average degree of the nearest neighbors of a node of fitness $x$  as:
\begin{equation}
k_{nn}(x)=\int dx'\, k(x')\phi(x'|x),
\end{equation}
where $\phi(x'|x)$ is the conditional probability that a node of fitness $x$ is connected to a node of fitness $x'$:
\begin{equation}\label{eq_phip}
\phi(x'|x)= \frac{\rho(x')\,f(x,x')}{\int {dx^{''}}\rho(x^{''})\,f(x,x^{''})} = \frac{N \rho(x') f(x,x')}{k(x)}
\end{equation}
that it the probability of finding a node of fitness $x'$ times the probability of creating such link.
Thus, one has that
\begin{equation}
k_{nn}(x)= \frac{N}{k(x)}\int dx'\, \rho(x')k(x')f(x,x')
\end{equation}
so the correlation function of average degree of nearest neighbor (ANND) is:
\begin{equation}\label{eq_ANND}
k_{nn}(k)= 1+\frac{1}{P(k)}\int dx \, \rho(x) p_L(k|x) k_{nn}(x) 
\end{equation}
where the propagator $p_L(k|x)$ has been found in our case to fulfill the poissonian distribution\footnote{ One can also define the probability that a node of actual degree $k$ has associated a fitness $x$ is called $p_L^*(x|k)$ that is the inverse of the propagator in the master equation eq.\eqref{eq:master_eq_p}, and it is defined via the Bayes' formula:
	\begin{equation*}
	P(k)\,p_L^*(x|k)=\rho (x)\,p_L(k|x)
	\end{equation*}} of eq.\eqref{eq:propaga_cont}.
Rewriting the eq.\eqref{eq_phip} we can write for the out-degree case:
\begin{align*}
k_{nn}(x) = & \frac{N}{k_0 x^{\alpha}}\int dx'\, k_0x'^{\alpha} \, Cx'^{-\mu} \, \frac{x^{\alpha}x'^{\beta}}{A^{\alpha +\beta}}\\
=& \frac{NC}{A^{\alpha +\beta}}\int_{a}^{b} x'^{\alpha+\beta-\mu}dx'\\
=& \frac{NC}{A^{\alpha +\beta}} \frac{x^{\alpha+\beta-\mu+1}}{\alpha+\beta-\mu+1}\Bigg |_{x=a}^{x=b}\\
=: & K_0
\end{align*}
which yields the ANND function to be:
\begin{align*}
\overline{k}_{nn}(k)=& 1+ \frac{K_0}{P(k)}\int dx \rho(x) p_L(k|x)\\
=& 1+K_0
\end{align*}
which is independent of $k$, a signature of lack of correlations.
The same is valid we choose the in-degree ANND choosing $\nu(x)=r_0 x^{\beta}$ and $k(x)=k_0 x^{\beta}$ .
Despite the fact that the ANND is not a exhaustive measure of  degree-correlations, it gives is  widely  used  to
measure dependencies between degrees of neighbor nodes in a network. 
It is worth to notice that this is a result of our specific choice of the linking function, in fact, other choices , for example a threshold function, brings to natural assortative or disassortative behaviors despite the fact that the degree distributions are scale free.
The assortativity mixing is expressed by the average nearest neighbor degree, computed as:
     \begin{equation}\label{eq_ANNDi}
		k_{nn,i}=\frac{1}{k_i}\sum_{j\in\varTheta (i)}k_j=\frac{1}{k_i}\sum_{j\neq i,1}^{N}A_{ij}k_j 
     \end{equation}    
where $k_i$ is the degree of node $i$ and $\varTheta (i)$ are the  neighbors of node $i$. Then the ANND function can be computed as the average value of eq.\eqref{eq_ANNDi} over all the node with a certain  degree $k$.
In the directed graph case, we split the ANND for both incoming and outgoing links as \citet{squartini2011analytical}:
\begin{align}
k_{nn,i}^{out}=& \; \frac{\sum_{j\neq i}\sum_{k\neq j} A_{ij}A_{jk}}{\sum_{j\neq i} A_{ij}}\\
k_{nn,i}^{in}=& \; \frac{\sum_{j\neq i}\sum_{k\neq j} A_{ji}A_{kj}}{\sum_{j\neq i} A_{ji}}
\end{align}
From the simulations over $100$ repetitions from Fig.\ref{fig:bidegree} we observe a disassortative behavior of our baseline network which is in contrast with the neutral mixing predicted by the analytical calculations. We will devote the next section to the discussion of this behavior in terms of finite-size effects.  
The fact that fitness model with factorizable linking function does not have any assortative behavior, derives from a derivation in which the system is closer to its more plausible configuration which correspond to run our incremental formulation  up to a time for which we get the average degree of the random version in the thermodynamic limit ($N \to \infty$). 
One of the most important applications of hidden variable model is that of creating network with a given degree correlation \citet{alessandro2007large}, and one of the simplest example is one for which a threshold linking function is chosen, as shown in \citet{boguna2003class}.
A comparison between the incremental procedure and the random version, can be useful to highlight the fact that there is a strong (apparent) discrepancy between the two generative prescriptions of the network. Since in the random model is not possible tho choose a given number of links (so the density), each simulation gives a different average degree which  fluctuates around the expected value (as calculated in appendix \ref{sec:scalefreebehavior1} for the case of truncated power law fitness distribution.). As consequences the ANND function shows a clear neutral assortativity as predicted by the theoretical calculations. As shown in Fig.\ref{fig_assortaKS} the incremental version allows to observe the assortative behavior for different replications with the same number of links. More the network is dense more the the ANND shows its spurious disassortativity due to the effect of finite size of the system. Anyway everything turns to coincide to the theoretical prediction for $N\to \infty$.

%\clearpage

\section{Conclusions}
In this paper we address some theoretical interesting features of scale free networks with an hidden variable structure, focusing on a incremental interpretation of the model. We gave a wave-like analytical description of the network dynamics which van be easily generalized for different choices of fitness distribution and pair linking function. Moreover this approach is  useful to stress the importance of  finite-size effects as the arise of correlations in not-infinite systems. As regard to some practical issues, the advantage of such procedure respect to the random generative procedure is the possibility to create a network with a desired number of links (i.e. average degree) in the network. This approach tunrs to be useful in simulations of contagion and cascades simulations in the theoretical and empirical stdies of financial networks where system with different degrees of inter-connectivity  is chosen to simulate different scenarios of systemic risks. The fact of having a incremental procedure allows also to save time in the computational selections of parameters for the contagion and cascade dynamics. Moreover we can understand how some selections of linking probability can give birth of spurious degree-correlations due to the finite size  of  the network. Ultimately a weighted version of the model is provided which solves the limitation of the theoretical description in a non sparse networks.

\section*{Acknowledgement}
Fabio Vanni acknowledges support from the European Union's Horizon 2020 research and innovation programme under grant agreement No.822781 GROWINPRO - Growth Welfare Innovation Productivity.
% % % % % % % % %       APPENDICI       % % % % % % % % % % % % % % % % % % % % % % % % % % % % % % % % % % % % % %

\newpage

\clearpage

\begin{appendices}
\begin{center}
\large\bfseries {\large APPENDICES}
\end{center}

 % % % % % % % % % % % % % % % % % % % % % % % % % % % % % % % % % % % % %5
 % % % % %$         solution    Master Equation      % % % % % % % % % % % % % % %
 %\newpage 
 \section{Solution to the Master Equation}\label{app_a}
 In the following section we describe three different ways to solve the master equation of the kinetic fitness network. In all the solutions the function $r(x)$ that depends on the fitness variable is not involved in the procedure until the integration for the total degree distribution. In this section we have to consider the couple of variables $(k,t)$ that is the degree of a node and the time of network evolution that express also the number of created links up the that time. We will use in sequential section the discrete-degree and discrete-time approach, then the discrete-degree and continuous-time approach, and finally the continuous-degree and continuous-time approach. 
 The original master equation for the evolution of the conditional-fitness probabilities is:
  \begin{align}\label{eq:master_eq_px}
 p_{t+1} (k|x) =&\;\big [1 - r(x)\big]\,p_{t} (k|x) + r(x)\, p_{t} (k-1|x),
  \end{align}
  taking in account the initial condition $p_{t=0}(k|x)=\delta_{0,k}$.
 
 \label{sec:mastereq}
 \subsection{Discrete  solution}\label{sec:mastereq_a}
 We consider discrete time step and discrete degree, as the most suitable case of network evolution.  Let us take discrete time steps 1 of $t=L$.
  Identifying the time with the number of links added: 
  %\begin{align}
  %& p_t(k|x)=  \big(1-r(x)\big)\,p_t(k|x) + r(x)\,p_t(k-1|x)\qquad \text{for } k\geq 1\\
  %& p_t(0|x) =  \big(1-r(x)\big)\,p_t(0|x) \nonumber
  %\end{align}
  \begin{equation}\label{eq:discreteME}
  \left\{
  \begin{array}{l l}
   p_t(k|x)=  \big(1-r(x)\big)\,p_t(k|x) + r(x)\,p_t(k-1|x)\qquad \text{for } k\geq 1\\
   p_t(0|x) =  \big(1-r(x)\big)\,p_t(0|x)  
  \end{array}
  \right.
  \end{equation}
  with initial condition $p_0(k|x)=\delta_{0,k}$.
  Considering the generating function for $k\geq 1$:
  \begin{equation*}
  G(s,k)=\sum_{t=0}^{\infty} s^t p_t(k|x)
  \end{equation*}
  multiplying both sides of eq.\eqref{eq:discreteME} by $s^t$ and summing over $t$:
  \begin{equation*}
  \sum_{t=0}^{\infty} s^{t} p_{t}(k|x)=(1-r(x))\sum_{t=0}^{\infty} s^{t} p_{t}(k|x) + r(x) \sum_{t=0}^{\infty} s^{t} p_{t}(k-1|x) 
  \end{equation*}
  the l.h.s is equivalent to:
  \begin{equation*}
  \frac{1}{s}\sum_{t=0}^{\infty} s^{t+1} p_{t+1}(k|x)= \frac{1}{s}\sum_{t=1}^{\infty} s^t p_t(k|x)= \frac{1}{s}(G(s,k)-p_0(k|x))
  \end{equation*}
  and the equation reduces to:
  \begin{equation*}
   \frac{1}{s}(G(s,k)-p_0(k|x)) = (1-r(x)) \, G(s,k) + r\,G(s,k-1).
  \end{equation*}
  Since  $p_0(k|x)=0$ and collecting common terms:
  \begin{equation*}
  G(s,k)= \frac{r(x)\,s}{1-s(1-r(x)) } G(s,k-1)
  \end{equation*}
  If $k=0$ it is 
  \begin{equation*}
  \frac{1}{s}(G(s,0)-p_0(0,x))= (1-r(x))G(s,0)
  \end{equation*}
  and since $p_0(0,x)=1$
  $$
  G(s,0)=\frac{1}{1-s(1-r(x))}
  $$
  so that in general
  \begin{equation}\label{eq:genefun}
  G(s,k)=\frac{s^k r(x)^k}{(1-s(1-r(x)))^{k+1}}
  \end{equation}
  Notice that 
  \begin{equation*}
  p_t(k,x)=\frac{1}{t!}\;\frac{d^t}{d\,s^t}G(s,k)\Big \rvert_{s=0}
  \end{equation*} 
  when taking the $t$ derivative of eq.\eqref{eq:genefun} computed in $s=0$, the result is zero if $t<k$. In general
  \begin{align}
  \frac{d^t}{d\,s^t}G(s,k) \Big \rvert_{s=0} = &  r(x)^k \sum_{n=0}^{t} {{t}\choose{n}} \frac{d^n}{d\,s^n}s^k\Bigg \rvert_{s=0} \nonumber \\  &\frac{d^{t-n}}{d\,s^{t-n}}\frac{1}{(1-s(1-r(x)))^{k+1}}\Bigg \rvert_{s=0} \nonumber
  \end{align}
  but the first term is different from zero only if $n=k$, then:
  \begin{align}
  p_t(k,x)=&\frac{1}{t!}\;\frac{d^t}{d\,s^t}G(s,k)\Big \rvert_{s=0} \\
  = & {{t}\choose{k}} \frac{r(x)^k k!}{t!} \frac{d^{t-k}}{d\,s^{t-k}}\frac{1}{(1-s(1-r(x)))^{k+1}}\Bigg \rvert_{s=0}    \nonumber \\
  = & {{t}\choose{k}}  \frac{ k!}{t!} r(x)^k (1-r(x))^{t-k}  \frac{ t!}{k!} \nonumber \\
  = & {{t}\choose{k}} r(x)^k\,(1-r(x))^{t-k} \label{eq:propaga_discrA}
  \end{align}
  Fixed at time $t$ the value of $k$ is distributed in $(0,\ldots , t)$ following a binomial distribution.
  
  Let us observe that for $t\to \infty$  the discrete Binomial distribution of eq.\eqref{eq:propaga_discr} tends to the  Poisson Distribution of eq.\eqref{eq:propaga_cont} in the regime $k \ll t$
 
 % % % % % % % % % % %
 
 \subsection{Continuous solution}\label{sec:mastereq_c}
  Let us now assume the we are exploring longer times and the system created a certain number of links. So passing to the continuous-time  with discretization step $\delta t = \Delta L = 1$ we can write:
  \begin{align}
  \frac{\partial p_{t}(k|x)}{\partial t} = & \; r(x)\big[   p_{t}(k-1|x) - p_{t}(k|x) \big]
  \end{align}
   The initial condition is that at time $t=0$ (where no link is present) the network is disconnected so that the nodes have degree $0$ and the conditional degree distribution is peaked at zero: $p_{L=0}(k|x)=\delta_{k,0}$.
   
  Using the Laplace transform to solve the differential equation we have:
  \begin{align}
  \mathcal{L}\big[ \dot{p}(k|x) \big] = s\,\mathcal{L}\big[ p(k|x)\big] = r(x) \mathcal{L}\big[ p(k-1|x)\big] -r(x) \mathcal{L}\big[ p(k|x)\big] 
  \end{align}  
  from which we can write :
  
  \begin{align}
  (s+r(x)) \mathcal{L}\big[ p(k|x)\big]= r(x) \mathcal{L}\big[ p(k-1|x)\big] \nonumber
  \end{align}
  
  and we can write the recursive equation:
  \begin{align}
  \mathcal{L}\big[ p(k|x)\big] =& \frac{r(x)}{s+r(x)} \mathcal{L}\big[ p(k-1|x)\big] \\
  =&  \left[ \frac{r(x)}{s+r(x)}\right]^k \mathcal{L}\big[ p(0|x)\big] 
  \end{align}
  
  from the initial condition we have that $\mathcal{L}\big[ p(0|x)\big] = \frac{1}{s+r(x)}$ we have :
   \begin{align}
   \mathcal{L}\big[ p(k|x)\big] =&  \frac{r(x)^k}{(s+r(x))^{k+1}}
   \end{align}
 
  the inverse Laplace transform give us the time evolution of the conditional probability:
  
  \begin{align}
  p_t(k|x)= \frac{[t\cdot \, r(x)]^k}{k!}e^{-t\cdot \, r(x)}\label{eq:propaga_contA}
  \end{align}
 
 % % % % % %        Evolution of P    % % % % % % % % % % % % % % % % % %

 \subsection{Degree Distribution}
 \label{sec:kineticapprox}
 \label{sec:kineticapprox1}
  Considering the case of a sparse condition, we have that $L\ll N(N-1)$ or equivently $\langle k \rangle \ll N-1$ or in a multi-links framework,  it is possible to write the expression (for the out-degree) as:
 \begin{equation}
 r(x)=r_0\,x^{\alpha},
 \end{equation}
 where the sparse condition  avoids  the dense condition of links in the network.
 
 Moreover, in the case of scale-free distribution 
 \begin{align}
 \rho(x)=& \rho_o\,x^{-\mu}
 \end{align}
 
 At this point the probability distribution becomes\footnote{where in our approach the time $t$ of the network evolution is equivalent to the number of created links $L$. In addition, we calculated the asymptotic behavior so neglecting the truncations and finitiness of the functions in the integral.} :
 \begin{align}
 P_L(k)=&\int_{0}^{\infty} p_L(k|x)\rho(x)dx, \\
 =&\, \frac{L^k}{k!}\int e^{-Lr(x)}\,r(x)^k\, \rho(x)\, dx, \\
 =& \,\rho _0 \frac{(Lr_0)^k}{k!}\int x^{k\alpha-\mu}\,e^{-Lr_0\,x^{\alpha}}\\
 =& \, \rho _0 \frac{(Lr_0)^k}{k!} \left [  \frac{(Lr_0)^{\frac{-1-k\alpha+\mu}{\alpha}}  \; \Gamma(\frac{1+k\alpha -\mu}{\alpha})}{\alpha}   \right]\\
 =& \, \frac{\rho _0}{\alpha} \, (Lr_0)^{\frac{\mu -1 }{\alpha}}\, \frac{\Gamma (k- \frac{\mu -1}{\alpha})}{\Gamma(k+1)} ,
 \end{align}
 under the condition of  $k> (\mu-1)/\alpha$.
 
An asyntotic expression can be obtained when $k$ is large (but far from the dense condition):
\begin{equation}
\frac{\Gamma (k- \frac{\mu -1}{\alpha})}{\Gamma(k+1)} \approx \left( \frac{1}{k} \right )^{\frac{-1+\mu}{\alpha}}\,\left( \frac{1}{k} + O\left(\frac{1}{k}\right)^2 \right)
\end{equation}

So the behavior of the (out)degree distribution for the sparse network is:
\begin{equation}\label{eq_degree_distrx}
P_L(k)\sim \, L^{\frac{\mu -1}{\alpha}}\, k^{-\frac{\alpha+\mu -1}{\alpha}} \qquad \text{for } \;   L = o(N^2)
\end{equation} 
this approximation becomes more precise when $N\to \infty$ .

% % % % % % % % % % % % % % % % % % % % % % % % % % %
% % % % %

 \section{Degree Transport equation: the asymptotic continuous solution}\label{sec:mastereq_d}
 We now add another description that is closer to a physical interpretation of transport through the  advection equation.
 Let us assume asymptotic behavior for the degree variable $k$ and the time $t$ so we can write:
 \begin{align}
p_x(k,t+1)-p_x(k,t)=&\; r_x [p_x(k,t)-p_x(k-1,t)]\\
\frac{\partial p_x(k,t)}{\partial t}+ r_x\frac{\partial p_x(k,t)}{\partial k}&=0 \qquad \text{for } \, t\to \infty  \, , \, k\to \infty \\
\Bigg( \frac{\partial }{\partial t}+ r_x\frac{\partial }{\partial k}\Bigg) p_x(k,t) &=0 \label{eq_advectionA}
 \end{align} 
 where we used the backward discrete differentiation for the degree and the forward differentiation for the time variable. 
 The equation represents the transport advection equation in one dimension, the solution can be found setting the initial condition as $p_x(k,t=0)= \delta_{k,0}$ where $\delta$ is the Kronecker symbol. At this point we assume (sparse network condition for $N\to \infty$) that it is a good approximation to let  $k \in (0,+\infty)$  and hence no need for boundary conditions.
 
 We can see each $p_x(k,t)$ for different fitnesses $x$ as wave packets of the transport equation each of those having a solution as $$p_x(k,t) = \phi_x(k-r_x t)=\delta (k-r_x t)$$  where $\phi_x$ is the kernel density which, in this case is  the impulse packet wave  function derived by the initial condition.
 The fitness-packet traveling waves proceed  with positive speed $r_x$, in a way that packets are  faster for larger fitness nodes and slower for smaller fitness nodes.

At this point the partial differential equation is a well-posed problem  and the general solution of the distribution $P(k,t)$ can be found as:
  \begin{equation}
P(k,t) = \int p_x(k,t)\rho(x)dx = \int \delta (k-r_x t)\rho(x)dx
  \end{equation} 
 where $\rho(x)$ is the nodes' fitness distribution and the integral limits depends on the choice of the fitness distribution. %and $x_0$ and $M$ the minimum and the maximum fitness respectively. 
 
 At this point we use the property of the Kronecker function:
 $$
 \delta(g(x))= \sum_{i=1}^{N}\frac{1}{|g'(x)|}\delta(x-s_i)
 $$
 where $s_i$ are the roots of $g(x)=0$.  In our case $g(x)=k-r_x t$ so the roots are $x_0 =\sqrt[\alpha]{\frac{k}{r_0 t}}$ since $r_x =r_0 x^{\alpha}$.
 
 As a consequence we get 
 $$g'(x)|_{x=x_0}=  -t r'(x)|_{x_0} = c_{\alpha} t^{\frac{1}{\alpha}}\, k^{\frac{\alpha -1}{\alpha }} $$
 where the constant $c_{\alpha}= \alpha r_0^{1-\frac{\alpha-1}{\alpha}}$
 At this point one can write the Kronecker function as:
 $$
 \delta(k- r_x t) = \frac{\delta \Big(x- \frac{k^{\frac{1}{\alpha}}}{ r_0^{\frac{1}{\alpha}} t^{\frac{1}{\alpha}} } \Big)}{c_{\alpha} t^{\frac{1}{\alpha}} k^{\frac{\alpha -1 }{\alpha}}}
 $$ 
 
 At this point we can write the solution for $N \to \infty$ of the degree distribution over time:
 \begin{align}
 P(k,t)=\; &  \frac{1}{c_{\alpha} t^{\frac{1}{\alpha}} k^{\frac{\alpha -1 }{\alpha}}}\int \delta \Bigg(x- \frac{k^{\frac{1}{\alpha}}}{ r_0^{\frac{1}{\alpha}} t^{\frac{1}{\alpha}} } \Bigg) \rho (x) dx \\
 =\; &  \frac{C}{c_{\alpha} t^{\frac{1}{\alpha}} k^{\frac{\alpha -1 }{\alpha}}}    \Bigg(  \frac{k^{\frac{1}{\alpha}}}{r_0^{\frac{1}{\alpha}} \,t^{\frac{1}{\alpha}}}  \Bigg)^{-\mu}\\
 = \;& \frac{C}{\alpha r_0^{\frac{1-\mu}{\alpha}}}\; t^{\frac{\mu-1}{\alpha}}\, k^{-\frac{\alpha +\mu-1}{\alpha}}
 \end{align} 
where we choose the Pareto-like distribution $\rho(x)=C x^{-\mu}$.
So, we recovered the same solution as in eq.\eqref{eq_degree_distrx} which in the truncated pareto fitness distribution $C= \frac{(\mu-1)a^{\mu -1}}{1-(\frac{a}{b})^{\mu -1}}$ where $x \in [a,b]$.

It is useful to stress that using this approach it is easy to calculate degree distributions for any choice of fitness distribution (as in Appendix \ref{sec:static_model} ) as long as the linking function is factorizable.

% % % % % % % % % % % % % % % % % % % % % % % % % % % % % % % %
\section{Bounded degree space: a discussion }\label{sec_bounded}
Until now all of the wave equations we have examined were considered on an semi-infinite domain $k\in(0 ,\infty )$ but to find the solution to eq.\eqref{eq_advectionA}, we need to specify the boundary conditions compatible with the finite number of neighbors per each node for a given fitness. 

Moreover, in principle we should consider a system of  partial differential equations per each fitness $x$, but we should consider that the fitness traveling packet $p_x$ does not evolve independently  since we should consider that each pair of nodes cannot create a link more than once.

Using a macroscopic approach, in the case of dense and finite network,  $L \approx N^2$, it is almost true that $k(x)\approx  constant $,  over nodes with different fitness $x$,in such situation from the definition eq.(\ref{eq:rate_edg_cont}) considering that 
\begin{align}
r(x)= \frac{k(x)}{N \int_0^{\infty} k(x)\rho (x)dx }.
\end{align}

we can write : 
\begin{equation*}
r(x)\approx \frac{1}{N}
\end{equation*}
so using the  binomial distribution for $p_L(k|z)$, we have;
\begin{align}
 P_L(k)\approx &\, {L \choose k} \left(\frac{1}{N}\right)^k \left(1-\frac{1}{N}\right)^{L-k} \quad \text{for }\, L=O(N^2) \nonumber
\end{align}
that is a  approximation starting from a time $t$ such that the various $k_i$ becomes to saturate to the common value.
Such binomial distribution has an average degree  that becomes closer and closer to a narrow peaked distribution for complete graphs. In the all-to-all situation  is the most dense network possible where $k_i=\langle k \rangle = N-1$.

The presence of this "hump effect" is also present for values of $L<N^2$ showing  an exponential decay in the power law tails of the survival function of the degree distributions. Moreover in the pdf function, a peak is always present in the tail as index of the condensation for the nodes with higher degrees. This  affects the exact slope of power law coefficient for values of $L$ closer and closer to dense condition of the network evolution.   

In conclusion, the case of $L=o(N^2)$ gives the possibility of a clear power law behavior in the degree distributions. In the situation where $L=O(N^2)$ the exponential behavior becomes more and more evident up to a  peaked function at maximum density of  full connected network.

In a microscopic view in terms of transport equations,  we should need to consider a bounded domain $ 0 < k < N $, and one can write the problem as:
\begin{align}
&\frac{\partial p_x(k,t)}{\partial t} +r_x\frac{\partial p_x(k,t)}{\partial k}  =0, \qquad \; 0<k<N \\
& p_x(k,0)= \delta(k), \frac{}{} \qquad \qquad \qquad \quad \; \; 0\leq k \leq N  \; \;  \; \; \; (\text{Initial Conditions}) \\
& p_x(0,t) = (1-r_x)^t,  \qquad \qquad \qquad \; t>0  \qquad \; \; \; \; \; \; \text{(Boundary Conditions})
\end{align}
which is properly posed in the sense that there is a unique solution which can be written  as (\citet{salsa2016partial}):
\begin{equation}
p_x(k,t) = \begin{cases} (1-r_x)^{t-k/r_x} &\mbox{if } \; k \leq r_x t\\
\delta(k-r_xt) & \mbox{if } \; k > r_x t \end{cases}
\end{equation}
where we have the relation:
\begin{align}
r_x= \frac{k(x)}{N \int k(x)\rho (x)dx }.
\end{align}
since $k(x)$ is the expected degree of a node with fitness $x$ as defined in eq.\eqref{eq:expected_degree1} we can write $r_x t = k(x)$  as long as $t=L$ i.e. $t\leq N(N-1)$.

Despite the previous  boundary-initial value problem for the transport equation is well posed and it has a unique e solution, it does not fully capture the dynamics (multilinks and self-loops) of the network and, moreover, it is not straightforward to derive  an analytical closed formulation of the degree distribution   $P(k,t)$. 

% the boundary condition for each wave packet is $p_x(k=N,t)=0$ that is sufficient to well define the problem since  this is a first-order equation.
%
%For finite networks two issues are very important, and they coincide with the same assumptions in the static solution of the fitness-models: we assume there are no multi-links and no self-loops. 
%
%  Moreover in a finite size network, the nodes with larger fitness are faster then low-fitness nodes at arriving at the boundary of maximum created links.
  
  An approximated solution can be attempted which takes in account looking at the problem as a transport equation with decay:
  \begin{equation}
   \frac{\partial }{\partial t}p_x(k,t)  + r_x\frac{\partial }{\partial k}p_x(k,t) = -\gamma \,p_x(k,t)
  \end{equation}
  where $\gamma$ is the decay which is proportional to the present density trough the relation $\gamma =  k(x)$. The meaning is that larger is the expected fitness degree  more intense is the effect of the damping. .
  
Using  the initial condition $p_x(k,0)=\delta(k)$, the solution of the problem takes the form of a   damped travelling fitness packet:
 \begin{equation}
 p_x(k,t)=\delta (k-r_xt)e^{-k(x) t}
 \end{equation}

At this point the (out)degree distribution as:
\begin{align}
 P(k,t)=\; &  \frac{1}{c_{\alpha} t^{\frac{1}{\alpha}} k^{\frac{\alpha -1 }{\alpha}}}\int \delta \Bigg(x- \frac{k^{\frac{1}{\alpha}}}{ r_0^{\frac{1}{\alpha}} t^{\frac{1}{\alpha}} } \Bigg) [e^{-k_0x^{\alpha} t} \rho (x)] dx \\
 \propto \;& t^{\frac{\mu-1}{\alpha}}\, k^{-\frac{\alpha +\mu-1}{\alpha}} e^{-\frac{k_0}{r_0}k}
 \end{align} 
 which is a power a power law function in $k$  with a exponential cutoff which can mimic the boundary effects of the degree space on the degree distribution.
 
%From the relation between $k(x)$ and $r(x)$, we can write  $k_0/r_0 = N\langle k \rangle $.
% \begin{align}
% r_0=\frac{N}{L}m_2(a,b,\mu; A, \alpha,\beta)
% \end{align}
% with $m_2$ is a constant (see eq.\eqref{eq:k_x}) depending on  the triple parameters $(a,b,\mu)$ (referred to the fitness distribution) and on the triple parameters $(A, \alpha,\beta)$ (referred to the linking function). If we take a truncated Pareto distribution for $\rho(x)$, we can make the choice $A=b$. 
% 
% In conclusion,  the decay exponent rate depends on all those parameters which are the expression of the finite-size effect. In particular in the case of  very sparse network $L\ll N$
%As regard with the number of nodes, in the most probable configuration of the static model, we have $L\propto  N^2$  and the decay rate is $r_0 \sim 1/N$ so that for an infinite network we have a negligible exponential truncation, as regard with the number of units. 

%However there are other factors that make the degree distribution truncated as $A$ and $b$ in the linking function and the fitness distribution respectively.
%
%As regard the kinetic prescription $L=t$ at early time, but $L\approx N^2$ near the dense network case. So that the truncation depends mainly from network parameters at early times and from $N$ for larger times as evident from Fig.\ref{fig:sparsedense}.
\newpage
% % % % % % % % % % % % % % % % % % % % % % % % % % % % % % %
 \section{Random  Generative Degree Distribution}\label{sec:static_model}
 The probabilistic fitness model describes a network that comprises a fixed number of nodes $N$ and edges $L$ . 
      
     The expected degree of a node with fitness $x$ is given by:
     \begin{equation}\label{eq:expected_degree}
     k(x)=N \int_{0}^{\infty} f(x,y)\rho(y)dy \equiv N\, F(x),
     \end{equation}
     and  the degree distribution can then be deduced by
     \begin{align}\label{eq:degree_kernel}
    P(k) = & \int \rho(x) \delta[k-k(x)]dx \nonumber \\
    = & \rho\left[ F^{-1}\left(\frac{k}{N}\right)  \right] \frac{d}{dk}F^{-1}\left(\frac{k}{N}\right),
     \end{align}
     where we have supposed that $F(x)$ is a monotonic function of $x$. From this last  equation, one can see that if the fitness distribution is a power law and $F$ is for  example linear the resulting network will be scale-free. This property, however,
     does not yield an explanation of the presence of scale-free degree distributions in  real complex networks since the use of a power-law $\rho(x)$ is not a priori justified.

     Let us now specify the particular choice of the linking function $f(x_i,x_j)$ and of the fitness distribution $\rho (x)$.

          The fitness distribution $\rho (x)$ should be normalized with a factor $C_{\rho}$ that  depends on the shape of power law function we use. Since the real world distribution are limited in size we use a truncated version of Pareto distribution in the interval $[a,b]$.
          
     The degree distribution  in the static network has a power law behavior  : 
         \begin{align}
         P(k_{in})= &  K_{\beta} k_{in}^{-\frac{\beta+\mu -1 }{\beta}}\\
         P(k_{out})= &  K_{\alpha} k_{out}^{-\frac{\alpha+\mu -1 }{\alpha}}
         \end{align}
     with specific proportionality coefficients as described in appendix \ref{sec:scalefreebehavior1}, \ref{sec:scalefreebehavior2} and \ref{sec:scalefreebehavior3} for different types of fitness function. We instead keep fixed the factorized choice of the linking function:
 \begin{equation}\label{eq:general_linking_ab}
        f(x_i,x_j)=  \frac{x_i^\alpha \, \cdot \, x_j^{\beta}}{A^{\alpha + \beta}},
 \end{equation}
 
   In order to consider the linking function as probability we should impose:
       \begin{equation}
       \int_{0}^{A}dx_i \int_{0}^{A}dx_j \; \frac{x_i ^{\alpha}x_j ^{\beta}}{A^{\alpha + \beta}} = 1 \nonumber
       \end{equation}
       Let us call $\xi := x_i /A $ and $\chi := x_j / A $, so we have:
       \begin{align}
       & f(\xi,\chi) = Z\; \xi^{\alpha}\; \chi ^{\beta} \qquad  \xi \in [0,1] , \chi \in [0,1] \\
       & Z=  \frac{1}{(\alpha+1) (\beta+1)} \qquad \alpha >-1, \beta >-1.
       \end{align}
       
       Such linking functions can create scale-free networks also for non scale-invariant fitness distributions $\rho (x)$, and then we will explore some of them.
       With this particular choice of the linking probability we can control various situations just playing with the coefficients $\alpha$ and $\beta$, creating different structure of networks.

 % % % % % % % %           TRUNCATED         % % % % % % % % % % % % % % % % % % % % % % %

 \subsection{ Truncated Power Law}\label{sec:scalefreebehavior1}
  
 In the truncated version of Pareto distribution $ \rho(x_i) = \bar{C}_{\rho} \,x_i^{-\mu},$ the constant factor becomes $$\bar{C}_{\rho}= \frac{(\mu-1)a^{\mu -1}}{1-(\frac{a}{b})^{\mu -1}}.$$ Notice that in this case $A=b$.
   
   The topological structure of the network of $N$ vertices is determined by the linking function and characterize the network in terms of degree distribution that can be obtain through the eq.\eqref{eq:expected_degree}:
   \begin{align}
   \frac{k_{in}(x)}{N} = F_{in}(x) =& \int_{a}^{b} f(s,x)\rho(s)ds \\
   = &\int_{a}^{b} \frac{s^{\alpha}x^{\beta}}{A^{\alpha + \beta}} \cdot \bar{C}_{\rho}\,s^{-\mu} ds \nonumber \\
   =& \frac{\bar{C}_{\rho} }{A^{\alpha + \beta}}x^{\beta}\int_{a}^{b} s^{-(\mu-\alpha)} ds    \nonumber \\
   =&   \frac{\bar{C}_{\rho} }{A^{\alpha + \beta}}x^{\beta} \;\left[ \frac{s ^{\alpha - \mu+1 } }{\alpha - \mu+1 }    \right]^b _a  \nonumber \\
   = & \left[  \frac{\bar{C}_{\rho} (b^{\alpha -\mu +1} - a^{\alpha -\mu +1}) }{A^{\alpha + \beta}(\alpha -\mu +1)}  \right ] \cdot x^{\beta}\label{eq_kxcontant},
   \end{align} 
   so it is possible to write:
   \begin{align}\label{eq:Fkx12}
   x=& \left[ \frac{ A^{\alpha + \beta}(\alpha -\mu +1) }{ N \bar{C}_{\rho} (b^{\alpha -\mu +1} - a^{\alpha -\mu +1}) }   \right]^{\frac{1}{\beta}} \, k_{in}^{\frac{1}{\beta}} \nonumber \\
   \equiv & \, M_1 \, k_{in}^{\frac{1}{\beta}}.
   \end{align}
   
   In order to get the in-degree distribution, let us use the eq.\eqref{eq:degree_kernel}, obtaining:
   \begin{align}
   P(k_{in})=& \left[  \bar{C}_{\rho}\left(M_1\, k_{in}^{\frac{1}{\beta}}\right)^{-\mu}  \right] \cdot \left[ \frac{M_1}{\beta}k_{in}^{\frac{1}{\beta}-1}\right ] \nonumber \\
   =& \frac{\bar{C}_{\rho} M_1^{1-\mu}}{\beta} \; k_{in}^{-\frac{\beta + \mu-1}{\beta}}  ,
   \end{align}
   we have a not divergent average degree only for the values $\beta\leq \mu-1$. Indeed 
   \begin{align}
   \langle k \rangle & =  \int_{k_{min}}^{k_{max}}dk\; k\,P(k) \nonumber \\
   & = \frac{\bar{C}_{\rho} M_1^{1-\mu}}{\beta} \frac{k_{max}^{2- \frac{(\beta + \mu-1)}{\beta}} - k_{min}^{2- \frac{(\beta + \mu-1)}{\beta}}}{\frac{(\beta - \mu +1)}{\beta}} \nonumber \\
   & = \bar{C}_{\rho} M_1^{1-\mu}  \;  \frac{k_{max}^{\frac{\beta - \mu+1}{\beta}} - k_{min}^{\frac{\beta - \mu+1}{\beta}}}{\beta - \mu +1} \quad \text{for $\beta \neq \mu-1$} \\
   & = \bar{C}_{\rho} M_1^{1-\mu}  \;  \log \frac{k_{max}}{k_{min}} \quad \text{for $\beta = \mu-1$} ,
   \end{align}
   then for $\beta > \mu -1$ if $k_{max}\to \infty$ the average degree diverges.
   As seen, for scale free networks (power law distribution) the largest value determines a natural cutoff and it follows:
   \begin{equation}
   k_{max}=k_{min}N^{\frac{1}{\mu -1}},
   \end{equation}
 so, when $N\to \infty$ we have an ideal power law behavior and the cut-off effect goes larger and larger ($k_{max}\to \infty$).
 %It is plausible that the contribution to $k_{min}$ come from the smallest fitness values $x \sim a $ so we can rewrite the average degree as:
 %\begin{align}
 %\langle k \rangle & = \frac{N}{A^{\alpha +\beta}}\, \frac{ a^{\alpha -\mu +1}\,   a^{\beta -\mu +1}}{(\alpha -\mu +1)(\beta -\mu +1)}
 %\end{align}
 %At this point let us calculated the average degree (in-degree and out-degree coincides as they should) in terms of fitness variable $x$. It is plausible that  substituting al the coefficient explicitly :
 %\begin{align}
 %\langle k \rangle & = \frac{N}{A^{\alpha +\beta}}\, \frac{(b^{\alpha -\mu +1} - a^{\alpha -\mu +1})\, (b^{\beta -\mu +1} - a^{\beta -\mu +1})}{(\alpha -\mu +1)(\beta -\mu +1)}
 %\end{align}
   
   For the out-degree probability we follow the same procedure:
     \begin{align}
     \frac{k_{out}(x)}{N} = F_{out}(x) =& \int_{a}^{b} f(x,s)\rho(s)ds \\
     = &\int_{a}^{b} \frac{x^{\alpha}s^{\beta}}{A^{\alpha + \beta}} \cdot \bar{C}_{\rho}\,s^{-\mu} ds \nonumber \\
     =& \frac{\bar{C}_{\rho} }{A^{\alpha + \beta}}x^{\alpha}\int_{a}^{b} s^{-(\mu-\beta)} ds    \nonumber \\
     =&   \frac{\bar{C}_{\rho} }{A^{\alpha + \beta}}x^{\alpha} \;\left[ \frac{s ^{\beta - \mu+1 } }{\beta - \mu+1 }    \right]^b _a  \nonumber \\
     = & \left[  \frac{\bar{C}_{\rho} (b^{\beta -\mu +1} - a^{\beta -\mu +1}) }{A^{\alpha + \beta}(\beta -\mu +1)}  \right ] \cdot x^{\alpha},
     \label{eq:k_x}
     \end{align} 
     from which:
       \begin{align}\label{eq:Fkx21}
       x=& \left[ \frac{ A^{\alpha + \beta}(\beta -\mu +1) }{ N \bar{C}_{\rho} (b^{\beta -\mu +1} - a^{\beta -\mu +1}) }   \right]^{\frac{1}{\alpha}} \, k_{out}^{\frac{1}{\alpha}} \nonumber \\
       \equiv & \, M_2 \, k_{out}^{\frac{1}{\alpha}}.
       \end{align}
       Finally we get the out-degree distribution as:
         \begin{align}
         P(k_{out})=& \left[  \bar{C}_{\rho} \left(M_2\, k_{out}^{\frac{1}{\alpha}}\right)^{-\mu}  \right] \cdot \left[ \frac{M_2}{\alpha}k_{out}^{\frac{1}{\alpha}-1}\right ] \nonumber \\
         =& \frac{\bar{C}_{\rho} M_2^{1-\mu}}{\alpha} \; k_{out}^{-\frac{\alpha + \mu-1}{\alpha}}  ,
         \end{align}
     and the average degree is not divergent for values $\alpha \leq \mu -1$.

     To summarize the previous calculation let us write:
     \begin{align}\label{eq_truncatedDegreeProb}
     P(k_{in})= &  K_{\beta} k_{in}^{-\frac{\beta+\mu -1 }{\beta}}\\
     P(k_{out})= &  K_{\alpha} k_{out}^{-\frac{\alpha+\mu -1 }{\alpha}}
     \end{align}

      In the first case it is possible to use the fitness distribution exponent $\mu$ ti tune the average degree.
      In the ideal case of truncated power law, we can write the average number of degree per node:
      
      \begin{align}
      \frac{\langle k_{in} \rangle}{N} =& \bar{C}_{\rho}  \left[  \frac{ (b^{\alpha -\mu +1} - a^{\alpha -\mu +1}) }{A^{\alpha + \beta}(\alpha -\mu +1)}  \right ] \cdot \langle x^{\beta}\rangle 
      \end{align} 
     but we have that :
     \begin{align}
      \langle x^{\beta}\rangle =& E(x^{\beta}) = \int_{a}^{b} x^{\beta} \bar{C}_{\rho}x^{-\mu} \nonumber \\
               =&  \bar{C}_{\rho} \frac{(b^{\beta -\mu +1} - a^{\beta -\mu +1}) }{\beta -\mu +1}
     \end{align}
     
     From these two equations we can finally write the average degree as:
     \begin{align}
     \langle k_{in} \rangle =& N\bar{C}_{\rho} ^2 \frac{ (b^{\alpha -\mu +1} - a^{\alpha -\mu +1})(b^{\beta -\mu +1} - a^{\beta -\mu +1})}{A^{\alpha+\beta}(\alpha -\mu +1)(\beta -\mu +1)}
     \end{align}
     where we have obviously $ \langle k_{in} \rangle = \langle k_{out} \rangle = \langle k \rangle $.

     % % % % % %     CUTOFF       % % % % % % % % % % % % % % % % % % % % % % % % % % % % % % % % % % % % % % % %
     \subsection{ Power law with exponential cutoff}\label{sec:scalefreebehavior2}
     In this case the fitness distribution $\rho (x)$ as a general power law with exponential cutoff in a real world situation where such distribution is defined in the interval  $[a,b]$:
            \begin{align}\label{eq:general_fitness}
            \rho(x_i) = & C_{[a,b]}\; x_i^{-\mu}\;e^{-\lambda x_i} \qquad \text{ with } \; x_i\in [a,b]\\
             = & \frac{\lambda ^{(1-\mu)}}{\Gamma(1-\mu,\lambda a)-\Gamma(1-\mu,\lambda b)} \,x_i^{-\mu}e^{-\lambda x_i},    
            \end{align} 
     where it is possible to recover the infinite cutoff power law  noticing that $$\lim\limits_{b\to \infty}\Gamma(1-\mu,\lambda b) = 0 $$     
     
      The degree distribution of the network  can be obtain through the eq.\eqref{eq:expected_degree}:
       \begin{align}
       \frac{k_{in}(x)}{N} = F_{in}(x) =& \int_{a}^{b} f(s,x)\rho(s)ds \\
       = &\int_{a}^{b} \frac{s^{\alpha}x^{\beta}}{A^{\alpha + \beta}} \cdot C_{[a,b]}\,s^{-\mu}\, e^{-\lambda s} ds \nonumber \\
       =& \frac{C_{[a,b]}}{A^{\alpha + \beta}}x^{\beta}\int_{a}^{b} s^{-(\mu-\alpha)}\,e^{-\lambda s} ds    \nonumber \\
       = &\frac{\lambda ^{-\alpha}}{A^{\alpha + \beta}} \left[  \frac{\Gamma (\alpha -\mu +1, \lambda a )-\Gamma (\alpha -\mu +1, \lambda b)}{\Gamma(1-\mu,\lambda a)-\Gamma(1-\mu,\lambda b)}   \right ] \cdot x^{\beta},
       \end{align} 
       where in the present case of expnential cutoff $b=A$.
       Finally,  it is possible to write:
       \begin{align}\label{eq:general_kx}
       x=&\left[ \frac{\lambda ^{\alpha} A^{\alpha + \beta}}{N} \frac{ \Gamma(1-\mu,\lambda a)-\Gamma(1-\mu,\lambda b)}{ \Gamma (\alpha -\mu +1, \lambda a)-\Gamma (\alpha -\mu +1,\lambda b) }   \right]^{\frac{1}{\beta}} \, k_{in}^{\frac{1}{\beta}} \nonumber \\
       \equiv & \, M_{\beta} \, k_{in}^{\frac{1}{\beta}}.
       \end{align}
       
       The in-degree distribution of the network becomes:
         \begin{align}
         P(k_{in})=& \left[C_{[a,b]} \left(M_{\beta}\, k_{in}^{\frac{1}{\beta}}\right)^{-\mu}\, e^{-\lambda \left(M_{\beta} k_{in}^{\frac{1}{\beta}}\right)}  \right] \cdot \left[ \frac{M_{\beta}}{\beta}k_{in}^{\frac{1}{\beta}-1}\right ], \nonumber \\
         =& \frac{C_{[a,b]} M_{\beta}^{1-\mu}}{\beta} \; k_{in}^{-\frac{\beta + \mu-1}{\beta}}e^{-\lambda M_{\beta} k_{in}^{\frac{1}{\beta}}}
         \end{align}
       
       As for the out-degree distribution we get similar results:
       \begin{align}
        \frac{k_{out}(x)}{N} &= \frac{\lambda ^{-\beta}}{A^{\alpha + \beta}} \left[  \frac{\Gamma (\beta -\mu +1, \lambda a )-\Gamma (\beta -\mu +1, \lambda b)}{\Gamma(1-\mu,\lambda a)-\Gamma(1-\mu,\lambda b)}   \right ] \cdot x^{\alpha}\\
         x&=\left[ \frac{\lambda ^{\beta} A^{\alpha + \beta}}{N} \frac{ \Gamma(1-\mu,\lambda a)-\Gamma(1-\mu,\lambda b)}{ \Gamma (\beta -\mu +1, \lambda a)-\Gamma (\beta -\mu +1,\lambda b) }   \right]^{\frac{1}{\alpha}} \, k_{out}^{\frac{1}{\alpha}} \nonumber \\
          & \equiv  \, M_{\alpha} \, k_{out}^{\frac{1}{\alpha}},\\
          P(k_{out})  &= \frac{C_{[a,b]} M_{\alpha}^{1-\mu}}{\alpha} \; k_{out}^{-\frac{\alpha + \mu-1}{\alpha}}e^{-\lambda M_{\alpha} k_{out}^{\frac{1}{\alpha}}}
       \end{align}
       
       If we set $\lambda =0$ we have the unbounded power law that is the Pareto distribution with $x_{min}=a$.
           For a detailed discussion on the truncated power law of this kind see \cite{bottazzi2009pareto}.
           
           % % % % % %     EXPONENTIAL       % % % % % % % % % % % % % % % % % % % % % % % % % % % % % % % % % % % % %

 \subsection{Exponential} \label{sec:scalefreebehavior3}
               Let us assume the case the fitness distribution $\rho (x)$ has an exponential behavior in the interval  $[a,b]$:
                 \begin{align}\label{eq:general_fitness}
                 \rho(x_i) = & C_{e}\;e^{-\lambda x_i} \qquad \text{ with } \; x_i\in [a,b]\\
                  = & \frac{\lambda}{e^{-\lambda a}- e^{-\lambda b}} \; e^{-\lambda x_i},    
                 \end{align} 
                 
                 and the expression for in-degree (out-degree) function is:
                 \begin{align}
                 x=&\left[\frac{A^{\alpha +\beta}\lambda^{1-\alpha}}{N\Gamma(1-\alpha,\lambda a)}\right]^{\frac{1}{\beta}} \; k^{\frac{1}{\beta}}\\
                 =& E_{\beta}\;k^{\frac{1}{\beta}}
                 \end{align}
                 and so the expression for the in-degree distribution is:
                 \begin{align}
                 P(k_{in})=&\frac{C_e}{\lambda \beta}\left(\lambda E_{\beta} \; k^{\frac{1}{\beta}-1}\right)\; e^{-\lambda E_{\beta}k^{\frac{1}{\beta}}}\\
                  P(k_{out})=&\frac{C_e}{\lambda \alpha}\left(\lambda E_{\alpha} \; k^{\frac{1}{\alpha}-1}\right)\; e^{-\lambda E_{\alpha}k^{\frac{1}{\alpha}}}
                 \end{align}
                 which are stretched exponential only if $\alpha >1$ and $\beta >1$. So, in this condition it is possible to hae a scale free distribution also for an exponential fitness density.
                 
\subsection{Lognormal}  \label{sec:scalefreebehavior4}
In the case the fitness distribution is  lognormal in the bounded support $(a,b)$  with two parameters $\mu $and   $\sigma$ that are, respectively, the mean and standard deviation of the variable’s natural logarithm:
     
                  \begin{align}\label{eq:lognormal_fitness}
                                  \rho(x) &= C_L \frac{1}{x\sigma \sqrt{2\pi}}\;e^{-\frac{(\log x -\mu)^2}{2\sigma ^2}}   \qquad \text{ with } \; x\in [a,b]
                   \end{align} 
                   where $C_L$ is the proper normalization constant so that $\int_{a}^{b}\rho(x)dx =1$.
                
  \begin{align}
      \frac{k_{out}(x)}{N} = F_{out}(x) =& \int_{a}^{b} f(x,s)\rho(s)ds \\
      = &\int_{a}^{b} \frac{x^{\alpha}s^{\beta}}{A^{\alpha + \beta}} \cdot  C_L \frac{1}{s\sigma \sqrt{2\pi}}\;e^{-\frac{(\log s -\mu)^2}{2\sigma ^2}} ds \nonumber \\
      =& \frac{C_L}{A^{\alpha + \beta}}x^{\alpha}\int_{a}^{b}\frac{s^{\beta -1}}{\sigma \sqrt{2\pi}}\;e^{-\frac{(\log s -\mu)^2}{2\sigma ^2}} ds \nonumber \\
      =&   \frac{C_L}{A^{\alpha + \beta}}x^{\alpha} \;\left[   \frac{2 \sigma ^2 e^{\frac{\mu }{2 \sigma ^2}} s^{\beta -\frac{1}{2 \sigma ^2}}}{2 \beta  \sigma ^2-1}  \right]^b _a  \nonumber 
      \label{eq:k_xLG}
      \end{align} 
      from which:
        \begin{align}
        x=& \, M_L \, k_{out}^{\frac{1}{\alpha}}.
        \end{align}    \label{eq:Fkx21_LG}           
 where $M_L $ is a shortcut coefficient depending on $N,\beta,\mu,\sigma$, so the expression for out-degree (in-degree) function is derived from:

    \begin{align}
         P(k_{out})=& \left[  C_L \frac{1}{(M_L  k_{out}^{\frac{1}{\alpha}} )\sigma \sqrt{2\pi}}\;e^{-\frac{(\frac{1}{\alpha} \log  k_{out} +\log M_L -\mu)^2}{2\sigma ^2}}\right] \cdot \left[ \frac{M_L}{\alpha}k_{out}^{\frac{1}{\alpha}-1}\right ] \nonumber \\
         =  &  \;   \frac{C_L}{\alpha \sigma \sqrt{2\pi}}\;k_{out}^{-1}\; e^{-(\frac{1}{\alpha} \log  k_{out} + \log M_L -\mu)^2/2\sigma ^2}
         \end{align}            
                   
                   which is a sort of modified stretched exponential function with a power law tail behavior.

 \subsection{Log Truncated Pareto}  \label{sec:scalefreebehavior5}                  
Let us study the case that the fitness variable is the logarithm             of a variable $s$ which follows a truncated Pareto      distribution in interval $[a,b]$ such that $x=g(s)=\log s$. 

Using the Jabobian transformation $ds/dx = e^x$ so that $\rho(x)=\rho(g^{-1}(x))|ds/dx|$, so that we can write the distribution of the logarithm of a truncated variable as:
\begin{align}
\rho(x)= & \;\bar{C}\; e^{-(\mu -1)x}\\
=&\; \frac{(\mu -1)a^{\mu-1}}{1-\left( a/b  \right)^{\mu-1}}\; e^{-(\mu -1)x}
\end{align} 
which we call Log-Pareto distribution\footnote{As regard with the moments of the such distribution, we can write:\begin{equation*}
E[x^k] = \int_{\log a}^{\log b} A x^k e^{-(\mu-1)x}dx
\end{equation*}
and in particular, the first two moments are:
 \begin{equation*}
 \begin{dcases*}
    E[x]= \frac{1}{(1-(a/b)^{\mu-1})} \left[    \log a +\frac{1}{\mu-1} - (a/b)^{\mu-1} \left(   \log b + \frac{1}{\mu-1} \right) \right]\\
    E[x^2] =  \frac{1}{(\mu-1)(1-(a/b)^{\mu-1})} \left[ \left[ (\mu-1)\log a +2 \right] \log a +\frac{2}{\mu-1}       -   \left[  \left( (\mu-1)\log b +2 \right)\log b +2  \right] (a/b)^{\mu-1}       \right]
 \end{dcases*}
 \end{equation*}
 
% In the unbounded version $b\to  \infty$, the the variance $Var(x) = E[x^2]-  \left(   E[x]\right)^2 $ is :
% \begin{align*}
% Var(x) \sim & \frac{  \left[  (\mu-1)\log a +2 \right]\log a  +\frac{2}{\mu -1}}{\mu-1} - \left( \log a +\frac{1}{\mu-1}  \right)^2 \qquad \text{for} \; \; b\gg a  
% \end{align*}
}.
A this point:
\begin{align*}
k_{out} = & \frac{N a^{\mu-1}}{A^{\alpha +\beta} (1-(a/b)^{\mu-1})} \Big[  -(\mu-1)^{-\beta-1}\Gamma(\beta+1) \, ,\, (\mu-1)x  \Big]_{x=a}^{x=b} \; x^{\alpha}
\end{align*}
so that $$x=m_{\beta} k^{\frac{1}{\alpha}}$$

finally the degree distributions are:
\begin{align}
P(k_{out}) = & \;\frac{\bar{C} m_{\beta}}{\alpha} \;  e^{-m_{\beta}(\mu-1)\;k_{out}^{1/\alpha}} \; k_{out}^{\frac{1}{\alpha} -1}\\
P(k_{in}) = &  \;\frac{\bar{C} m_{\alpha}}{\beta} \;  e^{-m_{\alpha}(\mu-1)\;k_{in}^{1/\beta}} \; k_{in}^{\frac{1}{\beta} -1}
\end{align}
 which are Weibull distributions which shows fat tails for $\alpha>1, \beta>1$ where the distribution is  subexponential  and it is considered heavy- tailed (see \citet{foss2013heavy}).
 
% which according to the value we assign to the parameters $\alpha , \beta$ it is possible to observe a power-law behavior with a stretched exponential cut-off, only if $a>1/2,b>1/2$

                 % % % % % % % % % % % % % % % % % % % % %
                 
\newpage
                 
 \section{Scale free regimes}\label{sec_regimes}
   Whatever we consider in-degree or out-degree distribution, they have the power law behavior of:
   \begin{equation}
   P(k) \propto  k^{-\tau},
   \end{equation}
     The fitness distribution density has a power coefficient such that $\mu > 1$.
 
 Considering a finite network with degree bounded in the interval $(k_{min},k_{max})$, the $m$th order moment is given by:
 
 \begin{equation*}
  \langle k ^m \rangle = \int_{k_{min}}^{k_{max}} k^mP(k)dk = 
 \begin{dcases*}
    \frac{C}{m+1-\tau}\left( k_{max}^{m+1-\tau}  - k_{min}^{m+1-\tau}\right)                       & if $\tau \neq m+1$ \\
   C \log(k_{max}/k_{min})& if $\tau = m+1$
 \end{dcases*}
 \end{equation*}

As regard to the first two moments  we have

 \begin{align}
  \langle k  \rangle &=
 \begin{dcases*}
    k_{max} ^{2-\tau}                  & if $\tau<2$ \\
   \log(k_{max})                          & if $\tau = 2$\\
   \text{finite}						     & if $\tau >2$
 \end{dcases*} \\
   \langle k ^2 \rangle  &= 
   \begin{dcases*}
      k_{max} ^{3-\tau}                  & if $\tau<3$ \\
     \log(k_{max})                          & if $\tau = 3$\\
     \text{finite}						     & if $\tau >3$
   \end{dcases*} 
 \end{align}
So the moment diverges as $k_{max}\to \infty$, growing with the upper degree.
An approximation value for the upper degree can be found considering that in scale free networks the number of nodes of degree greater than $k_{kmax}$ so that $N\int_{k_{max}}^{\infty}P(k)dk \simeq 1$, and so:
$$
k_{max}\simeq k_{min}N^{1/(\tau -1)}.
$$

 In the case of factorizable  $f(x,y)$ and pareto-fitness distribution $\rho(x)$,  the out or in degree distributions have the forms of  $$   P(k) \propto k^{-\frac{\gamma+\mu-1}{\gamma}} $$ 
 where $\gamma$ is the generic coefficient in the linking function, either $\alpha$ or $ \beta $ according to the fact that we are  considering the out- and the in- degree distribution respectively.
 
 We can summarize the scale-free regimes, with different regimes according to  linking and fitness the parameters $(\gamma, \mu)$ as: 
   \begin{description}
     \item[regular $2<\tau \leq 3$ .] Most of the real-world scale free networks live in this regime where the degree distributions have  finite first moments and infinite variances. In our case this requires a condition on $\mu$ and $\gamma$, that are:
     \begin{equation}
    \frac{\mu -1}{2}\leq\gamma < \mu -1
     \end{equation}
    \item[anomalous $1 <\tau \leq 2$ .]   In some case  scale-free real networks are in this regime in which even the first moment is infinite. The condition on the parameters $\mu$ and $\gamma$ are:
    \begin{equation}
    \gamma \geq \mu -1
    \end{equation}
    
    \item[random-like $\tau > 3$ .]   In this case the scaling coefficient of degree distributions has a value of $\tau $'s where scale-free networks are indistinguishable from a random network in the sense that in order  to document the scale free nature we should have a huge size of the network with an order of magnitude: $$N \sim \left(\frac{k_{max}}{k_{min}}\right)^{\tau-1 }.$$
    This regime happens when:
    \begin{equation}
    0<\gamma < \frac{\mu -1}{2}
    \end{equation}
     \end{description}
   
\begin{figure}[!ht]
    \centering
            \includegraphics[width=1.1\textwidth]{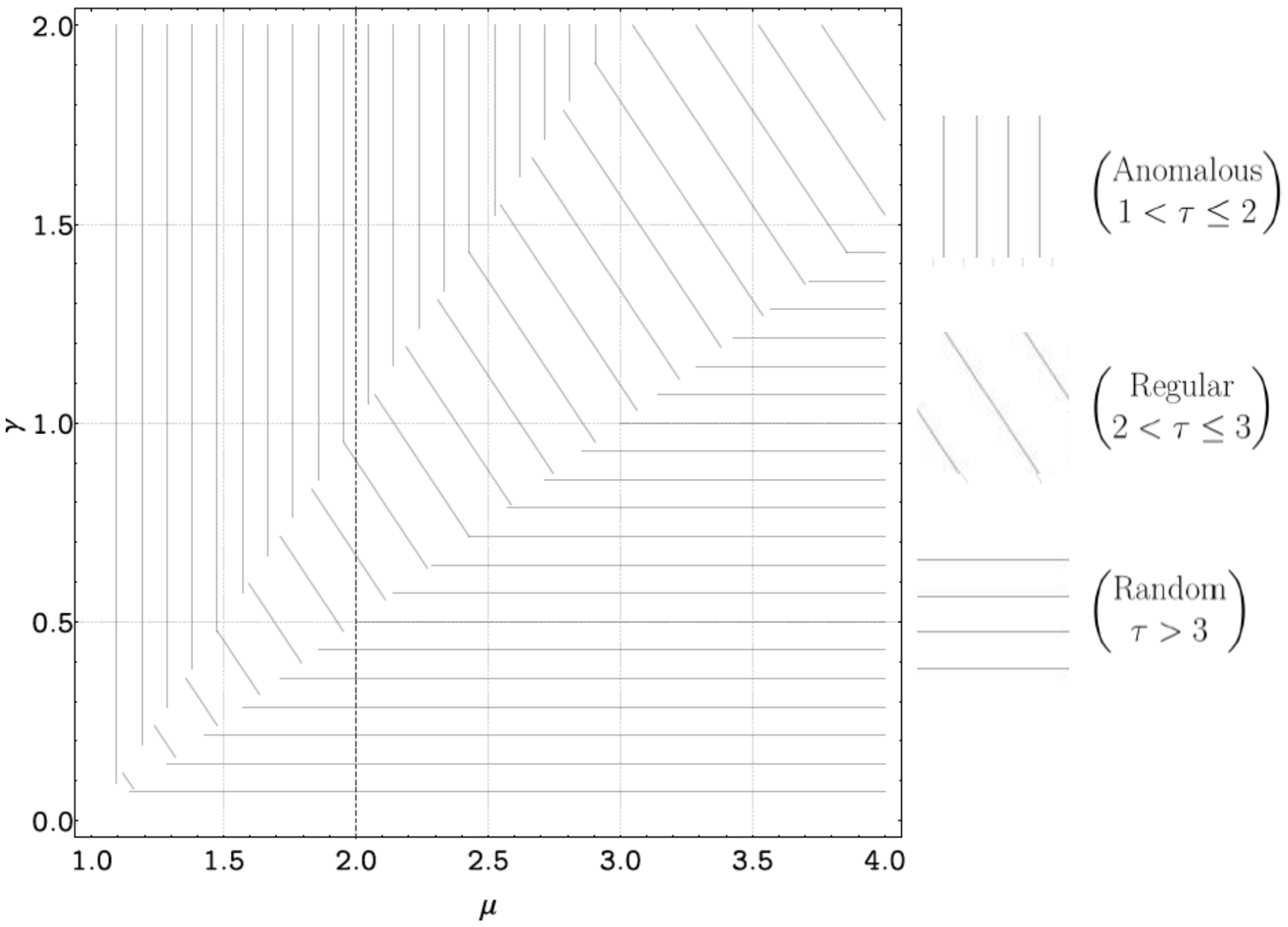}
            \caption{Region of the space of parameters $(\mu,\gamma)$ where a scale free network exhibits three different behaviors in their degree distributions. Clockwise, on the top we have anomalous regime, then in the middle, regular networks, and on the bottom random-like networks. The vertical dashed  black  line represents the classical case in which $\mu =2$ ( Lux \& Montagna). In this condition for the fitness density, regular scale free occurs for $\gamma$ in $(\frac{1}{2} , 1]$ .  The anomalous regime occurs when $\gamma > 1 $. Eventually, random-like behavior occurs for $ \gamma \in (0, \frac{1}{2}]$ }
            \label{fig:scale_free_regime}         
   \end{figure}

   All of these cases are summarized and showed in fig.\ref{fig:scale_free_regime}.
   
 This discussion can set a range for the out-degree and in-degree  coefficients choosing correct values $\alpha$ and $\beta$ of the linking function to have a scale free behavior in the degree-distributions.
 
 An important subcase of this network is for $$\gamma=\alpha = \beta =0$$ for which  $$ f(x_i,x_j)=const = c$$.
                    Then from the eq.\eqref{eq:expected_degree1}, using the power law distribution for the fitness function, we have:
                      \begin{align}
                      \frac{k_{out}(x)}{N} =& \int_{a}^{b} f(x,s)\rho(s)ds \\
                      = &\, c\int_{a}^{b}   C_{[a,b]}\,s^{-\mu} ds =c
                      \end{align} 
                      
                      From which we have the same degree per node, that is the case of the random graph with average degree controlled by the number of links created in the preferential attachment procedure:
                      \begin{equation}
                      \langle k\rangle =cN =  \frac{L}{N},
                      \end{equation}
                      from which $ c=L/N^2$ which plays the role of probability in Erdos-Renyi network.
                 
\begin{figure}[!ht]
      \centering
     	 \begin{subfigure}[c]{0.3\textwidth}
                       \includegraphics[width=\linewidth]{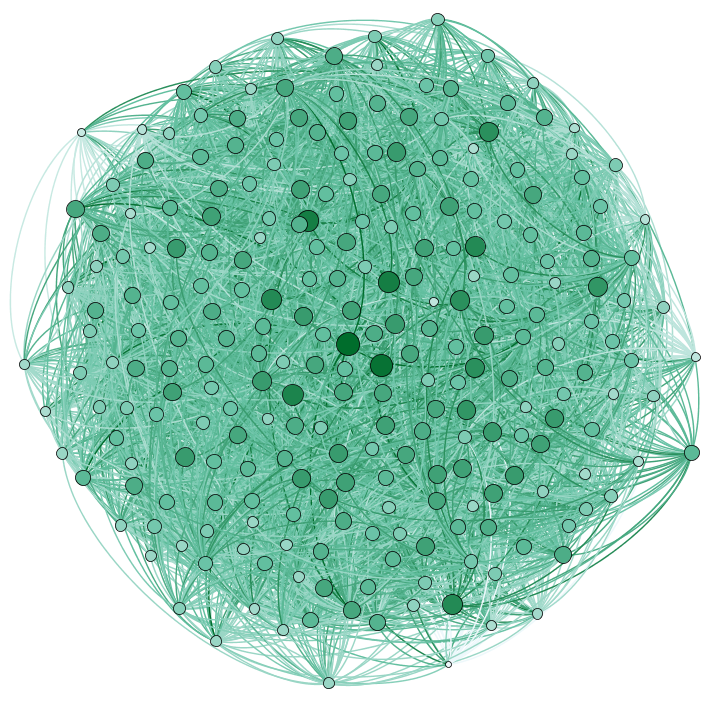}
                       \caption{Structure of bank network in the  regime of random  graph}
                       \label{fig:regime_r}
              \end{subfigure}%
      	         \qquad%\vspace{0.25\textwidth}
      	 \begin{subfigure}[c]{0.35\textwidth}
      	                  \includegraphics[width=\linewidth]{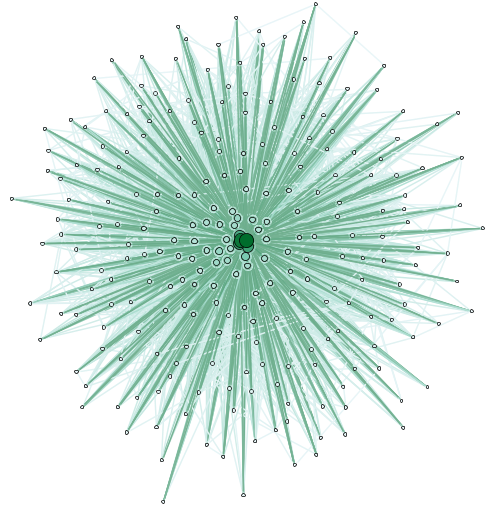}
      	                  \caption{Scale free bank network in the  regime of  graph where it is evident the core-periphery attitude in such kind of networks.}
      	                  \label{fig:regime_sf}
               \end{subfigure}%
               \caption{Layout of random (a) and scale free (b) networks.}\label{fig:layout}
              %add desired spacing between images, e. g. ~, \quad, \qquad etc. (or a blank line to force the subfigure onto a new line)
      \end{figure}

                 In this case non scale free behavior exists.  
                 A big difference between a random-like network and a scale-free one is also evident in the network layout as in fig.\ref{fig:layout}, where the scale-free network shows a core-periphery structure.
 
%\input{appendice_fowerfit.tex}
%\newpage
%\input{appendice_histpower.tex}
%\include{appendix}
\end{appendices}
\clearpage
%\appendix\label{sec:appendix}
%\include{appendix}
%\include{discussedresults}

\newpage
%\theendnotes

%\addcontentsline{toc}{section}{Bibliography} 
%\bibliographystyle{plain} 
\bibliographystyle{apalike} 
\bibliography{bibliografia_financial_networks}

\end{document}